\documentclass[12pt,preprint,showpacs,preprintnumbers]{revtex4}
\usepackage{amsfonts}
\usepackage{amsmath}
\usepackage{amssymb}
\usepackage{hyperref}
\usepackage{color}
\usepackage{graphicx}
\usepackage{amssymb}
\usepackage{amsmath}
\usepackage{graphicx}
\usepackage{dcolumn}
\usepackage{bm}
\usepackage{epsfig}
\usepackage[T1]{fontenc}
\usepackage{ae,aecompl}

\begin{document}

\title{Lattice Boltzmann modeling and simulation of compressible flows}
\author{ Aiguo Xu$^{1}$\footnote{%
Corresponding author. Email address: Xu\_Aiguo@iapcm.ac.cn}, Guangcai Zhang$%
^{1}$, Yanbiao Gan$^{2}$, Feng Chen$^{3}$, Xijun Yu$^{1}$}
\affiliation{$^1$ National Key Laboratory of Computational Physics, \\
Institute of Applied Physics and Computational Mathematics, P. O. Box
8009-26, Beijing 100088, P.R.China \\
$^2$ North China Institute of Aerospace Engineering, Langfang 065000,
P.R.China \\
$^3$ School of science, Linyi University, Linyi 276005, P. R. China}
\date{\today }

\begin{abstract}
In this mini-review we summarize the progress of Lattice Boltzmann(LB)
modeling and simulating compressible flows in our group in recent years.
Main contents include (i) Single-Relaxation-Time(SRT) LB model supplemented
by additional viscosity, (ii) Multiple-Relaxation-Time(MRT) LB model, and
(iii) LB study on hydrodynamic instabilities. The former two belong to
improvements of physical modeling and the third belongs to simulation or
application. The SRT-LB model supplemented by additional viscosity keeps the
original framework of Lattice Bhatnagar-Gross-Krook (LBGK). So, it is easier
and more convenient for previous SRT-LB users. The MRT-LB is a completely
new framework for physical modeling. It significantly extends the range of
LB applications. The cost is longer computational time. The developed SRT-LB
and MRT-LB are complementary from the sides of convenience and applicability.
\end{abstract}

\maketitle

\section{Introduction}

During the past two decades the lattice Boltzmann (LB) method has emerged as
a competitive scheme for simulating various nearly incompressible complex
flows \cite{Succi-Book}, ranging from magnetohydrodynamics  \cite%
{Chensy-PRL-1991,CICP-2008}, to flows of suspensions  \cite{suspension-1},
flows with phase separation \cite{chromodynamic model, shanchen,Yeomans, HCZ
model,XGL1,XGL2,XGL3,Succi-PRE-2007,Sofonea-multiphase1,Sofonea-multiphase2}%
, flows through porous media \cite{Succi-epl-1989-porous media,Xu-CTP-2008},
etc. With increasing the Mach number, the compressibility of flow becomes
more pronounced. Such high speed compressible flows are ubiquitous in
explosion physics, aerophysics and astrophysics, etc. Up to now, the LB
modeling and simulating of compressible flows, especially those with shocks
and/or discontinuities, is still a challenging issue.

Given the great importance of shocking and detonation in many fields of physics and
engineering \cite{Detonation,WangC1,WangC2}, constructing LB models for high speed
compressible flows has been attempted since the early days of LB research \cite%
{Succi-Book}. To proceed, we first discuss the most fundamental problem \textquotedblleft what is LB ?\textquotedblright.
The views are not exactly the same in
papers by different authors. Since having different knowledge backgrounds
and working in different fields, different authors may use LB to solve
different problems and focus on different sides of LB. Understandably, even
for the same author, the views will be updated with extending research
experience. Globally speaking, the views on LB can be classified into two
categories. The first category regards LB as a new scheme for simulating
hydrodynamic equations such as the Euler equations and Navier-Stokes
equations. The second category regards LB as a kind of new model of physical
systems. Physical model construction and numerical method design are the first two steps for
numerical study on any physical problems. Compared with numerical methods,
the physical model construction is the first step and more fundamental.
Only after the physical model is fixed can the corresponding numerical method be established.
Clearly,
the first kind of view starts LB research from the second step, numerical
method design. It does not consider the improvements of the physical modeling. In
other words, it assumes that the original hydrodynamic equations are
sufficiently exact for modeling the problem under consideration. The second
kind of view puts LB research on the more fundamental step, physical
modeling. For this view the numerical method is the second important issue.
It accepts any reasonable numerical methods no matter they are new or
traditional. The second kind of view aims at physical problems. The point of
the second view is that, compared with the traditional hydrodynamic
equations, the LB framework contains more physical components. The
theoretical reasons are as below. The LB model is based on the Boltzmann
equation which is one of the most fundamental equations in non-equilibrium
statistical physics. It naturally inherits some intrinsic characteristics of
the latter. According to the Chapman-Enskog analysis, one can expand the
distribution function around its equilibrium as Taylor series in the Knudsen
number. When the Knudsen number approaches zero,  the
system is nearly in equilibrium state, the deviation from equilibrium is
negligible, the LB model corresponds to or recovers the Euler equations.
When the first order terms in Knudsen number have to be accounted and the
second order terms are negligible, in other words, when the system 
slightly deviates from the equilibrium, the LB model corresponds to or
recovers the Navier-Stokes equations. When the system deviates more from
equilibrium and the second order terms in Knudsen number have to be taken
into account, the LB model is beyond the Navier-Stokes description. The
theoretical framework of LB is self-adaptive for describing complex systems
where the deviations from equilibrium are spatially and temporally varying.
From the view of modeling precision on detailed dynamics, it is less than
Molecular Dynamics(MD). It adopts the concept of distribution function. It
is generally considered as a kind of mesoscopic modeling. For continuum system,
the LB should give the same results as those of hydrodynamics equations. For
non-continuum systems such as the boundary layers where the Knudsen number
is high, the LB should give the same results as those of other mature
methods such as MD or Monte Carlo(MC). In between the two kinds of limiting
cases, the hydrodynamic equations are not valid, the MD and MC are
reasonable but not practical due to the huge quantity of computations. For
such cases, the LB modeling and simulation still work. Its results should be
checked by physical principles and analyses. Just as in traditional
Computational Fluid Dynamics(CFD) where different discretization schemes
work for different problems, for different systems one should compose or
choose different LB models.

In 1992 Alexander et al \cite{Alexander1992} proposed a compressible LB model
where the main skill is to introduce a flexible sound speed so that the Mach
number may become higher. This model works only for nearly isothermal
compressible systems. In 1999 Yan et al \cite{Yan1999} proposed a LB scheme
for compressible Euler equations. In this model a Discrete Velocity
Model(DVM) with three energy levels is used. Sun et al \cite{Sun1998,Sun2003}
proposed an adaptive LB model where the particle velocities vary with the
Mach number and internal energy. The model partly frees the particle velocity
from fixed values. It works for more extensive systems compared with
previous LB versions. Its two-dimensional and three-dimensional versions
were published in 1998 and 2003, respectively. The evolutions of all those
models follow the traditional \textquotedblleft propagation +
collision\textquotedblright\ mode. All of them belong to the standard LB
models. Due to the inconvenience of application and/or numerical instability
problems, few physical results based on those models can be found.

For modeling and simulating compressible flows, an alternative way is to use
the Finite-Difference(FD)-LB method. Tsutahara group \cite%
{Kataoka2004a,Kataoka2004b,Watari2003,Watari2004,Watari2007} in Kobe
university proposed several FD-LB models in recent years. The FD-LB model
frees the combination of spatial and temporal discretizations. The sizes of
particle velocities are flexible. So it is much more convenient to meet
the requirements for simulating compressible fluids. The FD-LB scheme was
then extended to the case of binary fluids \cite{Xu2005EPL,Xu2005PRE}. But
numerical instability problem blocks its practical applications to systems
with a Mach number being larger than $1$. In fact, as for the numerical
instability problem, many attempts have been made. Typical examples are
referred to the entropy LB model \cite{Karlin2002,Karlin2003}, FIX-UP scheme%
 \cite{Li2004}, flux-limters approach\cite{SLG2004}, etc. But most of the discussions
were still focused on systems with small Mach numbers.

To model and simulate high speed compressible flows, especially those with
shocks, our group developed two schemes in recent years. The first is to
introduce additional viscosity and improve the discretization of spatial
and temporal derivatives \cite%
{XuPan2007,XuGan2008PhysA,XuGan2008CTP,XuChen2009,XuChen2010,XuGan2011CTP}.
This scheme does not change the framework of the original LB model. The
second is to develop Multiple Relaxation Time(MRT) LB models \cite%
{MRT2011PLA,MRT2010EPL,MRT2011CTP,XuChen2011,MRT2011TAML}. The framework is
changed in the second scheme. The first scheme is based on the following
facts. (i) The numerical fluid particles do not distinguish the original
viscosity and additional viscosity. (ii) Introducing additional viscosity is
equivalent to modifying the relaxation time from some sense. (iii) Better
template of discretization may damp the numerical anisotropy. Our improved
models work for both high speed and low speed flows. So, they make it
possible to simulate stable shocks in compressible fluids. The first scheme
is based on the original Bhatnagar-Gross-Krook(BGK) model. It is a remedy
under the original framework.

The rest of the paper is structured as follows. We first introduce a few
improved LB models based on the first scheme in section II. The MRT scheme
is reviewed in section III. Section IV shows two typical applications, LB
studies on Richtmyer-Meshkov(RM) and Kelvin-Helmhotz(KH) instabilities.
Section V summarizes the present paper.

\section{SRT model supplemented by additional viscosity}

Among the two-dimensional FD-LB models for compressible flows, the one by
Kataoka and Tsutahara \cite{Kataoka2004a} is typical. It has very simple and
strict theoretical background, uses a DVM with only
$9$ components. The specific heat ratio is flexible. But the numerical
instability blocks its application in supersonic flows. Therefore, our first
LB model for high speed compressible flows is created by improving the
Kataoka-Tsutahara(KT) model.

The LB kinetic equation with BGK approximation reads,

\begin{equation}
\frac{\partial {{f}_{i}}}{\partial t}+{{v}_{i\alpha }}\frac{\partial {{f}_{i}%
}}{\partial {{x}_{\alpha }}}=\frac{1}{\tau }[{{f}_{i}}^{eq}-{{f}_{i}}]
\label{e1}
\end{equation}%
where $f_{i}$ ($f_{i}^{eq}$) is the discrete (equilibrium) distribution
function; $\mathbf{v}_{i}$ is the $i$-th discrete velocity, $i=0$, $\cdots$,
$N-1$; $N$ is the total number of the discrete velocity; index $\alpha =1$, $%
2$, $3$ corresponding to $x$, $y$, and $z$, respectively; $\tau $ is the
relaxation time determining the speed of approaching equilibrium. Sometimes,
$\tau $ is rewritten as $\epsilon \tau ^{\prime }$, where $\epsilon $ is a
dimensionless number, the Knudsen number. The original KT model corresponds
to the complete Euler equations

\begin{eqnarray}
\frac{\partial \rho }{\partial t}+\frac{\partial (\rho u_{\alpha })}{%
\partial x_{\alpha }} &=&0\text{,}  \notag \\
\frac{\partial (\rho u_{\alpha })}{\partial t}+\frac{\partial (\rho
u_{\alpha }u_{\beta })}{\partial x_{\beta }}+\frac{\partial P}{\partial
x_{\alpha }} &=&0\text{,}  \label{e2} \\
\frac{\partial }{\partial t}(E+\frac{1}{2}\rho u_{\alpha }^{2})+\frac{%
\partial }{\partial x_{\alpha }}[u_{\alpha }(E+\frac{1}{2}\rho u_{\beta
}^{2}+P)] &=&0\text{,}  \notag
\end{eqnarray}%
when the knudsen number $\epsilon $ approaching zero. Here $\rho $, $u$, $P$
($=\rho T$), $E$($=\rho T/(\gamma -1)$) are the hydrodynamic density, flow
velocity, pressure and internal energy, respectively; $T$ is the temperature
and $\gamma $ is the specific-heat ratio. To make $\gamma $ flexible, a
constant, $b=2/(\gamma -1)$, is introduced. The following constraints are
needed for this model,
\begin{equation}
\rho =\sum_{i=0}^{N-1}f_{i}^{eq}=\sum_{i=0}^{N-1}f_{i}\text{,}  \label{e4}
\end{equation}%
\begin{equation}
\rho u_{\alpha }=\sum_{i=0}^{N-1}f_{i}^{eq}v_{i\alpha
}=\sum_{i=0}^{N-1}f_{i}v_{i\alpha }\text{,}  \label{e5}
\end{equation}%
\begin{equation}
\rho (bRT+u_{\alpha }^{2})=\sum_{i=0}^{N-1}f_{i}^{eq}(v_{i\alpha }^{2}+\eta
_{i}^{2})=\sum_{i=0}^{N-1}f_{i}(v_{i\alpha }^{2}+\eta _{i}^{2})\text{,}
\label{e7}
\end{equation}%
\begin{equation}
P\delta _{\alpha \beta }+\rho u_{\alpha }u_{\beta
}=\sum_{i=0}^{N-1}f_{i}^{eq}v_{i\alpha }v_{i\beta }\text{,}  \label{e6}
\end{equation}%
\begin{equation}
\rho \left[(b+2)RT+u_{\beta }^{2}\right] u_{\alpha
}=\sum_{i=0}^{N-1}f_{i}^{eq}(v_{i\beta }^{2}+\eta _{i}^{2})v_{i\alpha }\text{%
,}  \label{e8}
\end{equation}%
where $\eta _{i}$ is another variable introduced to make specific-heat ratio
flexible.

In the two-dimensional case, the KT DVM has nine components. It reads
\begin{equation}
\begin{split}
(v_{i1},v_{i2})= \text{\hspace{3cm}}\\
\left\{
\begin{array}{ll}
(0,0), & i=0  \\
c_{1}[\cos (\frac{\pi (i+1)}{2}),\sin (\frac{\pi (i+1)}{2})]\text{,} &
i=1,2,3,4 \\
c_{2}[\cos \pi (\frac{i+1}{2}+\frac{1}{4}),\sin \pi (\frac{i+1}{2}+\frac{1}{4%
})]\text{,} & i=5,6,7,8%
\end{array}%
\right.   \label{e14b}
\end{split}
\end{equation}

\begin{equation}
\eta _{i}=\left\{
\begin{array}{ll}
\eta _{0}\text{,} & i=0 \\
0\text{,} & i=1,2,...,8%
\end{array}%
\right. .  \label{e15}
\end{equation}

A schematic figure of the distribution of the discrete velocities is shown
in Fig.1, where $c_{1}$ and $c_{2}$ are constants which should not depart
faraway from the flow velocity $u$. $c_{2}$ is generally chosen $1.0\sim 3.0$
times of $c_{1}$.

\begin{figure}[tbp]
\includegraphics*[ width=0.25\textwidth]{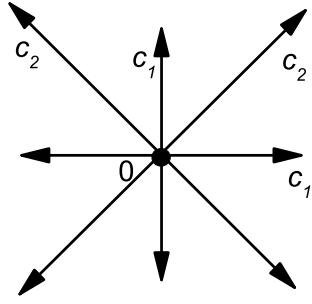}
\caption{Schematic figure of the discrete velocity model.}
\end{figure}

The local equilibrium distribution function is computed by
\begin{equation}
f_{i}^{eq}=\rho (A_{i}+B_{i}v_{i\alpha }u_{\alpha }+D_{i}u_{\alpha
}v_{i\alpha }u_{\beta }v_{i\beta })\text{, }i=0\text{,}1\text{,}
\cdots \text{,}8\text{,}  \label{e16}
\end{equation}%
where
\begin{widetext}
\begin{equation}
A_{i}=
\left\{
\begin{array}{ll}
\frac{b-2}{\eta _{0}^{2}}T\text{,} & \text{ }i=0 \\
\frac{1}{4(c_{1}^{2}-c_{2}^{2})}\left[ -c_{2}^{2}+\left( (b-2)\frac{c_{2}^{2}%
}{\eta _{0}^{2}}+2\right) T+\frac{c_{2}^{2}}{c_{1}^{2}}u_{\alpha }^{2}\right]
\text{, } & i=1,2,3,4 \\
\frac{1}{4(c_{2}^{2}-c_{1}^{2})}\left[ -c_{1}^{2}+\left( (b-2)\frac{c_{1}^{2}%
}{\eta _{0}^{2}}+2\right) T+\frac{c_{1}^{2}}{c_{2}^{2}}u_{\alpha }^{2}\right]
\text{,} & \text{ }i=5,6,7,8%
\end{array}%
\right.   \label{e17}
\end{equation}

\begin{equation}
B_{i}=\left\{
\begin{array}{ll}
0, & \quad i=0   \\
\frac{-c_{2}^{2}+(b+2)T+u_{\beta }^{2}}{2c_{1}^{2}(c_{1}^{2}-c_{2}^{2})}%
,& \quad i=1,2,3,4   \\
\frac{-c_{1}^{2}+(b+2)T+u_{\beta }^{2}}{2c_{2}^{2}(c_{2}^{2}-c_{1}^{2})}%
,& \quad i=5,6,7,8
\end{array}%
\right. \text{, }D_{i}=\left\{
\begin{array}{ll}
0,& \quad i=0  \\
\frac{1}{2c_{1}^{4}},& \quad i=1,2,3,4  \\
\frac{1}{2c_{2}^{4}},& \quad i=5,6,7,8
\end{array}%
\right.  \label{e18}
\end{equation}%

Parameters $\eta _{0}$, $c_{1}$ and $c_{2}$ are independent in this DVM. $%
\eta _{0}$ influences $f_{i}^{eq}$ via the expansion coefficient $A_{i}$. In
the original KT model, the usual FD scheme with first-order forward in time
and second-order upwinding in space is used.

To make practical the LB simulation to the supersonic flows, we propose an
alternative FD scheme combined with an additional dissipation term to
overcome the numerical instability problem. The LB equation \eqref{e1} can
be regarded as non-dimensional. In this work, we consider $\tau =\epsilon
\tau ^{\prime }$ and set the time step $\Delta t$ to be numerically equal to
the Knudsen number $\varepsilon $. Thus, from Eq.\eqref{e1} we have
\begin{equation}
\begin{split}
f_{i}(\mathbf{x},t+\Delta t)-f_{i}(\mathbf{x},t)+v_{i\alpha }\frac{\partial
f_{i}(\mathbf{x},t)}{\partial x_{\alpha }}\Delta t
=\frac{1}{\tau }\left[
f_{i}^{eq}(\mathbf{x},t)-f_{i}(\mathbf{x},t)\right] \text{.}
\label{SEC-VON-1}
\end{split}
\end{equation}%

In Eq.\eqref{SEC-VON-1} $\tau ^{\prime }$ has been written as $\tau $ for
simplicity. The spatial derivative $\partial f_{i}/\partial x$ can be
calculated by
\begin{equation}
\text{If }v_{ix}\geq 0\text{,}\quad \frac{\partial {f_{i}}}{\partial x}=%
\frac{\beta f_{i}(x+\Delta x,t)+(1-2\beta )f_{i}(x,t)-(1-\beta
)f_{i}(x-\Delta x,t)}{\Delta x}\text{;}  \label{SEC-VON-2a}
\end{equation}%
\begin{equation}
\text{If }v_{ix}<0\text{,}\quad \frac{\partial {f_{i}}}{\partial x}=\frac{%
(1-\beta )f_{i}(x+\Delta x,t)-(1-2\beta )f_{i}(x,t)-\beta f_{i}(x-\Delta x,t)%
}{\Delta x}\text{.}  \label{SEC-VON-2b}
\end{equation}%
In Eqs.\eqref{SEC-VON-2a} and \eqref{SEC-VON-2b}, $0\leq \beta \leq 0.5$. If
$\beta $ takes zero, then they are no other than the first-order upwind
scheme in space; if $\beta $ takes $0.5$, they recover to the general
central difference scheme. $\partial f_{i}/\partial y$ can be calculated in
a similar way. Actually, Eqs.\eqref{SEC-VON-2a} and \eqref{SEC-VON-2b} can
be rewritten as
\begin{eqnarray}
\text{If }v_{ix} &\geq &0\text{,}\quad \frac{\partial {f_{i}}}{\partial x}=%
\frac{f_{i}(x,t)-f_{i}(x-\Delta x,t)}{\Delta x}  \label{SEC-VON-2c} \\
&&+\frac{\beta \Delta x[f_{i}(x+\Delta x,t)+f_{i}(x-\Delta x,t)-2f_{i}(x,t)]%
}{{\Delta x}^{2}}\text{;}  \notag
\end{eqnarray}%
\begin{eqnarray}
\text{If }v_{ix} &<&0\text{,}\quad \frac{\partial {f_{i}}}{\partial x}=\frac{%
f_{i}(x+\Delta x,t)-f_{i}(x,t)}{\Delta x}  \label{SEC-VON-2d} \\
&&-\frac{\beta \Delta x[f_{i}(x+\Delta x,t)+f_{i}(x-\Delta x,t)-2f_{i}(x,t)]%
}{{\Delta x}^{2}}\text{.}  \notag
\end{eqnarray}%
The second terms in the Right-Hand-Side(RHS) of Eqs.\eqref{SEC-VON-2c} and %
\eqref{SEC-VON-2d} can be regarded as some kind of additional viscosities
which can reduce some unphysical phenomena such as wall-heating, but they
are not enough. Additional dissipation term is needed. The final LB equation
reads
\begin{equation}
f_{i}(\mathbf{x},t+\Delta t)-f_{i}(\mathbf{x},t)+v_{i\alpha }\frac{\partial
f_{i}(\mathbf{x},t)}{\partial x_{\alpha }}\Delta t-\lambda _{i}\sum_{\alpha
=1}^{2}\frac{\partial ^{2}f_{i}(\mathbf{x},t)}{\partial x_{\alpha }^{2}}%
\Delta t=\frac{1}{\tau }[f_{i}^{eq}(\mathbf{x},t)-f_{i}(\mathbf{x},t)]
\label{SEC-VON-11}
\end{equation}%
\end{widetext}
where $\lambda _{i}$ is a small number not varying in space or time. The
second-order derivative $\frac{\partial ^{2}f_{i}(\mathbf{x},t)}{\partial
x_{\alpha }^{2}}$ can be calculated by the central difference scheme. In our
simulations $\Delta x=\Delta y$ and the parameter $\beta $ is generally
chosen to be $0.25$ if not particularly claimed. How to choose the $\lambda
_{i}$ is the key problem. Analysis by the software, Mathematica, and
numerical tests show that we can choose $\lambda _{i}$ around the following
way,
\begin{equation}
\lambda _{i}=\left\{
\begin{array}{ll}
c_{1}\Delta x, & i=0 \\
c_{1}\Delta x/10, & i=1,2,3,4 \\
0, & i=5,6,7,8%
\end{array}%
\right. \text{.}  \label{SEC-VON-13}
\end{equation}

The improved model is validated by well-known benchmark tests. Simulations
on Riemann problems with very high ratios (1000:1) of pressure and density
also show good accuracy and stability. Regular and double Mach shock
reflections are successfully simulated. It should be commented that, since
using constraint, $\Delta t = \epsilon$, such a model can only be regarded
as a new scheme to simulate the Euler equations. The added viscosity terms
can be regarded as a kind of slight remedy to the traditional hydrodynamic
model.

\begin{figure}[tbp]
\includegraphics*[ width=0.48\textwidth]{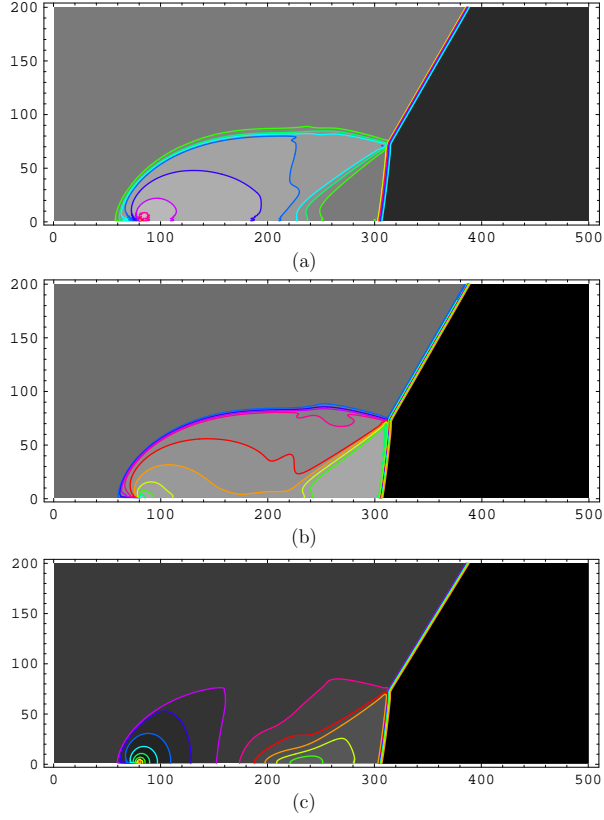}
\caption{Contours of density (a), temperature (b), and $u_{x}$ (c) of the
double Mach reflection problem at the time $t = 7.5\times 10^{-3}$. The
units of the $x$- and $y$- axes are both $0.001$. }
\end{figure}

In 2008 Gan, Xu, Zhang, et al \cite{XuGan2008PhysA} developed a LB model for
high speed compressible flows. In this model, the constraint, $\Delta
t=\epsilon $, is eliminated. Therefore, it can be regarded as a mesoscopic
new model. In the continuum limit it corresponds to the Navier-Stokes
equations. The model is composed of three components: (i) the DVM by Watari
and Tsutahara \cite{Watari2003}, (ii) a modified Lax--Wendroff FD scheme
where reasonable dissipation and dispersion are naturally included, (iii)
additional viscosity. The improved model is convenient to compromise the
high accuracy and stability. The included dispersion term can effectively
reduce the numerical oscillation at discontinuity. Shock tubes and shock
reflections are used to validate the new scheme. In our numerical tests the
Mach numbers are successfully increased up to 20 or higher. In Fig.2 we show
a simulation result on double Mach reflection by the improved model. The
initial pressure ratio here is high. A planar shock is incident towards an
oblique surface with a $30^{\circ }$ angle to the direction of propagation
of the shock. A uniform mesh size of $500\times 200$ is used for the
numerical simulation. The conditions for both sides are:
\begin{equation}
\begin{split}
\left( \rho ,u_{x},u_{y},T\right) \mid _{x\text{,}y\text{,}0}= \text{\hspace{3cm}} \\
\left\{
\begin{array}{ll}
(\frac{400}{67}\text{,}13.3\cos 30^{\circ }\text{,}-13.3\sin 30^{\circ }%
\text{,}89.2775)\text{,} & \text{ if }y\geq h(x\text{,}0) \\
(2.0\text{,}0.0\text{,}0.0\text{,}0.5)\text{,} & \text{ if }y<h(x\text{,}0)%
\end{array}%
\right. \text{,}
\end{split}
\end{equation}%
where $h(x,t)=\sqrt{3}(x-80\Delta x)-40t$. The reflecting wall lines along
the bottom of the problem domain, beginning at $x=0.08.$ The shock makes a $%
60^{\circ }$ angle with the $x$ axis and extends to the top of the problem
domain at $y=0.2$. At the top boundary, the physical quantities are assigned
the same values as on the left side for $x\leq g(t)$ and are assigned the
same values as on the right side for $x > g(t)$, where $g(t)=80\Delta x+\sqrt{3}/3(0.2+40t)$%
. The computed density, temperature and flow velocity along the $x$%
-direction are shown in Fig.2, where complex characteristics, such as
oblique shocks and triple points, are well captured.

In this model the ratio of specific heat is fixed on an unphysical constant
2. Later, Gan, Xu, Zhang, et al studied a model for flexible specific heat
ratio \cite{XuGan2008CTP}. For higher computational efficiency, Chen, Xu,
Zhang, et al proposed a model where the number of discrete velocity
decreases from 65 to 16 \cite{XuChen2009}. They simulated the reaction of
shock wave on a bubble or ball, etc. In 2010 they present a
three-dimensional LB model for high Mach number compressible flows. Figures
3(a) and 3(b) show our successful LB simulations of shock wave reactions
on bubble and on ball, respectively, where only the density isosurfaces are
shown. In both Figs.(a) and (b), the upper plot shows the initial state, and
the lower one shows a snapshot in the shocking procedure. The added
additional viscosity makes the scheme more consistent with the physical
system and more convenient to satisfy the von Neumann stability condition.
Among the discussions on LB model with additional viscosity, the application
of flux limiters is also investigated \cite{XuGan2011CTP}. In the reference
with flux limiters \cite{XuGan2011CTP} Gan, Xu, Zhang, et al also introduced
an improved BGK model to break the fixed-Prandtl-number barrier. It is
meaningful to briefly review the scheme for this improvement.

In the SRT model, both the viscosity and heat conductivity coefficients are
proportional to the relaxation time $\tau $. As a result, the Pr is fixed to
\begin{equation}
\Pr =\frac{c_{p}\mu }{\kappa }=1\text{.}
\end{equation}%
The control of $\Pr $ may be achieved by modifying the BGK collision term as
below:
\begin{equation}
\frac{\partial f_{ki}}{\partial t}+\mathbf{v}_{ki}\cdot \frac{\partial f_{ki}%
}{\partial \mathbf{r}}=-\frac{1}{\tau }\left[ f_{ki}-(1+\Lambda \tau
)f_{ki}^{eq}\right] \text{,}  \label{LB-Iki}
\end{equation}%
where $\Lambda $ takes the following form
\begin{equation}
\Lambda =A+B(\mathbf{v}_{ki}-\mathbf{u})^{2}\text{.}  \label{Iki}
\end{equation}%
Contributions of the new term $\Lambda f_{ki}^{eq}$ in Eq.(\ref{LB-Iki}) to
the mass, momentum, and energy equations are
\begin{equation}
\sum_{ki}\Lambda f_{ki}^{eq}=(A+2BT)\rho =0\text{,}  \label{Iki-1}
\end{equation}%
\begin{equation}
\sum_{ki}\Lambda f_{ki}^{eq}v_{ki\alpha }=(A+2BT)\rho u_{\alpha }=0\text{,}
\label{Iki-2}
\end{equation}%
\begin{equation}
\sum_{ki}\frac{1}{2}\Lambda f_{ki}^{eq}v_{k}^{2}=\rho (A+2BT)(T+\frac{u^{2}}{%
2})+2\rho T^{2}B=2\rho T^{2}B\text{.}  \label{Iki-3}
\end{equation}%
We require that Eq.(\ref{LB-Iki}) recovers the Navier-Stokes equations in
the following form,
\begin{widetext}
\begin{equation}
\frac{\partial \rho }{\partial t}+\frac{\partial (\rho u_{\alpha })}{%
\partial r_{\alpha }}=0\text{,}  \label{NS-1}
\end{equation}%
\begin{equation}
\frac{\partial (\rho u_{\alpha })}{\partial t}+\frac{\partial (\rho
u_{\alpha }u_{\beta }+P\delta _{\alpha \beta })}{\partial r_{\beta }}-\frac{%
\partial }{\partial r_{\beta }}[\mu (\frac{\partial u_{\beta }}{\partial
r_{\alpha }}+\frac{\partial u_{\alpha }}{\partial r_{\beta }}-(\gamma -1)%
\frac{\partial u_{\gamma }}{\partial r_{\gamma }}\delta _{\alpha \beta })]=0%
\text{,}  \label{NS-2}
\end{equation}%
\begin{gather}
\frac{\partial }{\partial t}[(E+\frac{\rho u^{2}}{2})]+\frac{\partial }{%
\partial r_{\alpha }}[u_{\alpha }(E+\frac{\rho u^{2}}{2}+P)]-\frac{\partial
}{\partial r_{\alpha }}[\kappa \frac{\partial T}{\partial r_{\alpha }}
\notag \\
+\mu u_{\beta }(\frac{\partial u_{\beta }}{\partial r_{\alpha }}+\frac{%
\partial u_{\alpha }}{\partial r_{\beta }}-(\gamma -1)\frac{\partial
u_{\gamma }}{\partial r_{\gamma }}\delta _{\alpha \beta })]=0\text{,}
\label{NS-3}
\end{gather}%
\end{widetext}
where $\mu =\rho T\tau $ is the viscosity, $\kappa $ is the heat
conductivity. $\kappa $ is required to be $\kappa =c_{p}\rho T(\tau +q)$,
where $c_{p}=\gamma c_{v}=\gamma /(\gamma -1)$ is the specific-heat at
constant pressure. It is clear that a new coefficient $q$ is introduced to
make the Prandtl number flexible. By using Eqs.(\ref{Iki-1})-(\ref{Iki-3})
it is easy to find coefficients in Eq.(\ref{Iki}) with the following form
\begin{equation}
A=-2BT\text{, }B=\frac{1}{2\rho T^{2}}\partial _{\alpha }[c_{p}\rho
Tq\partial _{\alpha }T]\text{.}
\end{equation}%
Therefore, the modified BGK collision term changes the heat conductibility
in the energy equation from $\kappa =c_{p}\rho T\tau $ to $\kappa =c_{p}\rho
T(\tau +q)$. Consequently, the Prandtl number is changed to
\begin{equation}
\Pr =\frac{\tau }{\tau +q}\text{.}
\end{equation}

Figure 4 shows a validation example of such a scheme for flexible Prandtl
numbers based on the SRT model. The figure shows the comparison of LB
results with theoretical solutions for thermal Couette Flows. Fig.(a) is for
the temperature profiles in steady state for various Prandtl numbers.
Fig.(b) shows the velocity profiles for $\Pr =5.0$ at various times. For more
details the readers can refer to Ref. \cite{XuGan2011CTP}. Such a scheme makes a
significant remedy from the side of physical modeling. It is easy to find
that such a scheme can also be used to change other transport coefficients
such as the viscosity. It is also meaningful to mention that among the
moment relations required by each LB model, only for the three, the
definitions of density, momentum and energy, the equilibrium distribution
function $f_{i}^{eq}$ can be replaced by the distribution function $f_{i}$.
If we replace $f_{i}^{eq}$ by $f_{i}$ in RHS of any other required moment
relations, the value of RHS will have a deviation from that of the left hand
side. This deviation may work as a measure for the deviation of system from
its equilibrium. For example, the following $\Delta _{1}$
\begin{equation}
\Delta _{1}=\sum_{i=0}^{N-1}f_{i}v_{i\alpha }v^2_{i\beta
}-\sum_{i=0}^{N-1}f_{i}^{eq}v_{i\alpha }v^2_{i\beta }\text{,}
\end{equation}%
presents a measure for how much the system deviates from its equilibrium for
cases without using the constraint $\Delta t=\epsilon $.

\begin{figure}[tbp]
\center\includegraphics*
[ width=0.7\textwidth]{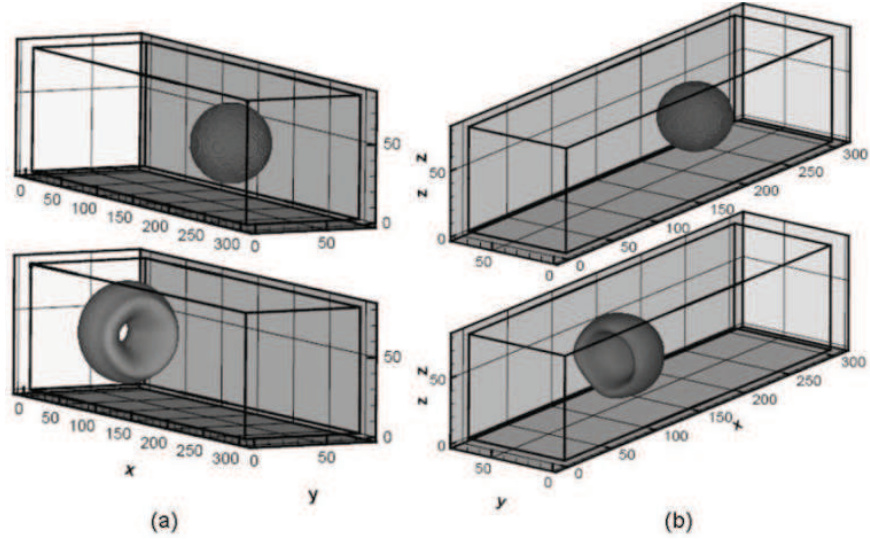}
\caption{Density isosurfaces of shocked bubble (a) and shocked ball (b). In
(a) or (b) the upper plot shows the initial state, the bottom one shows the
density configuration during the shocking procedure. }
\end{figure}

\begin{figure}[tbp]
\includegraphics*[ width=0.68\textwidth]{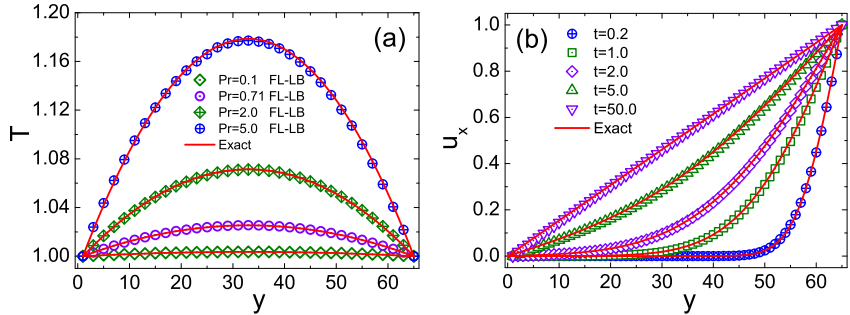}
\caption{ Comparison of LB results and theoretical solutions for thermal
Couette Flows. (a) Temperature profiles in steady state for various Prandtl
number. (b) Velocity profiles for $\Pr=5.0$ at various times.}
\end{figure}

\section{MRT model}

It is known that different motion modes generally approach their
equilibria in different velocities. But in the SRT BGK model, the speeds
of all discrete distribution functions approaching the equilibria are
determined by a single relaxation time $\tau $. That means $\tau $ is an
averaged relaxation time of all kinds of motion modes. The best merit of
this treatment is that it is simple and keeps the most fundamental
conservation laws. This BGK model has been successfully applied in various
fields. But with increasing the Mach number and Reynolds number, the problem
of numerical instability becomes more serious. At the same time, the Prandtl
number effect is a key issue in many fluidic systems. Facing with all these
requirements and challenges, people began to reevaluate this simple
averaging treatment.

The numerical instability of LB simulation is still a difficult problem
nowadays. Roughly speaking, the possible reasons come from two sides, the
physical modeling and the discretization scheme. It has been indicated that
untying the motion modes which should be independent is helpful for
improving the numerical stability \cite{SucciMRT1,SucciMRT2,Luo1,Luo2,Luo3}.
Succi, et al \cite{SucciMRT1}, Luo, et al \cite{Luo1,Luo2,Luo3} and many
others have made significant contributions in constructing MRT LB models.
Those MRT models are mainly within the framework of the standard LB model
and work for isothermal systems with low Mach number. In recent years our
group proposed two schemes to compose MRT model for high speed compressible
flows. These schemes are for the framework of the FD-LB model. The finished
works focus still on the two-dimensional cases.

In the MRT LB formulation, the collision step is first calculated in the
kinetic moment space spanned by a suitable set of $N$ kinetic moments of the
distribution function $f_{i}$. Then, the propagation step is performed back
in the discrete velocity space spanned by the $N$ discrete velocities $\mathbf{v%
}_{i}$. In contrast to the SRT model, the MRT version caters for more
adjustable parameters and degrees of freedom. The relaxation rates of the
various kinetic moments due to particle collisions may be adjusted
independently. The MRT LB equation has the following form,
\begin{equation}
\frac{\partial f_{i}}{\partial t}+v_{i\alpha }\frac{\partial f_{i}}{\partial
x_{\alpha }}=-\mathbf{S}_{ik}\left[ f_{k}-f_{k}^{eq}\right] \text{,}
\label{1}
\end{equation}%
where $\mathbf{S}$ is the collision matrix. The equation reduces to the
usual lattice BGK equation if all the relaxation parameters are set to be a
single relaxation time $\tau $, namely $\mathbf{S}=\frac{1}{\tau }\mathbf{I}$%
, where $\mathbf{I}$ is the identity matrix. The discrete distribution
functions $f_{i}$ and $f_{i}^{eq}$ can be rewritten as the following
matrixes:
\begin{subequations}
\begin{equation}
\mathbf{f}=\left( f_{1},f_{2},\cdots ,f_{N}\right) ^{T}\text{,}  \label{2a}
\end{equation}%
\begin{equation}
\mathbf{f}^{eq}=\left( f_{1}^{eq},f_{2}^{eq},\cdots ,f_{N}^{eq}\right) ^{T}%
\text{,}  \label{2b}
\end{equation}%
\end{subequations}
where $T$ is the transpose operator. Given a set of discrete velocities $%
\mathbf{v}_{i}$ and corresponding distribution functions $f_{i}$, we can get
a velocity space $S^{V}$ spanned by discrete velocities $\mathbf{v}_{i}$ and
a moment space $S^{M}$ spanned by moments of particle distribution function $%
f_{i}$. The moments of particle distribution function reads $\hat{\mathbf{f}}%
=\left( \hat{f}_{1},\hat{f}_{2},\cdots ,\hat{f}_{N}\right) ^{T}$, where $%
\hat{f}_{i}=m_{ij}f_{j}$, $m_{ij}$ is an element of the matrix $\mathbf{M}$
and is a polynomial of discrete velocities. Obviously, the moments are
simply linear combination of distribution functions$\ f_{i}$, and the
mapping between moment space and velocity space is defined by the linear
transformation $\mathbf{M}$, i.e., $\hat{\mathbf{f}}=\mathbf{Mf}$, $\mathbf{%
f=M}^{-1}\hat{\mathbf{f}}$, where $\mathbf{M}=\left( m_{1},m_{2},\cdots
,m_{N}\right) ^{T},m_{i}=(m_{i1},m_{i2},\cdots ,m_{iN})$.

Since the collision step is first calculated in the moment space and then
mapped back to the velocity space. So, the MRT LB equation can be described
as
\begin{equation}
\frac{\partial f_{i}}{\partial t}+v_{i\alpha }\frac{\partial f_{i}}{\partial
x_{\alpha }}=-\mathbf{M}_{il}^{-1}\hat{\mathbf{S}}_{lk}(\hat{f}_{k}-\hat{f}%
_{k}^{eq})\text{,}  \label{3}
\end{equation}%
where $\hat{\mathbf{S}}=\mathbf{MSM}^{-1}=diag(s_{1},s_{2},\cdots ,s_{N})$
is a diagonal relaxation matrix. $\hat{f}_{i}^{eq}$\ is the equilibrium
value of the moment $\hat{f}_{i}$. The moments can be divided into two
groups. The first group consists of the moments locally conserved in the
collision process, i.e. $\hat{f}_{i}=\hat{f}_{i}^{eq}$. The second group
consists of the moments not conserved, i.e. $\hat{f}_{i}\neq \hat{f}%
_{i}^{eq} $. The equilibrium $\hat{f}_{i}^{eq}$\ is a function of conserved
moments. It is clear that the first group includes the density, the momentum
and the energy.


\subsection{MRT model based on group representation theory}

Now we briefly review the first MRT LB model proposed in our group \cite%
{MRT2011PLA}. Our first MRT model is developed from the SRT version by
Kataoka and Tsutahara \cite{Kataoka2004b}. The DVM can be expressed as:
\begin{equation}
\left( v_{ix},v_{iy}\right) =\left\{
\begin{array}{cc}
\mathbf{cyc}:\left( \pm 1,0\right) \text{,} & \text{for }1\leq i\leq 4\text{,%
} \\
\mathbf{cyc}:\left( \pm 6,0\right) \text{,} & \text{for }5\leq i\leq 8\text{,%
} \\
\sqrt{2}\left( \pm 1,\pm 1\right) \text{,} & \text{for }9\leq i\leq 12\text{,%
} \\
\frac{3}{\sqrt{2}}\left( \pm 1,\pm 1\right) \text{,} & \text{for }13\leq
i\leq 16\text{,}%
\end{array}%
\right.  \label{4}
\end{equation}%
where \textbf{cyc} indicates the cyclic permutation. (see Fig. 5)%

\begin{figure}[tbp]
\center\includegraphics*[width=0.3\textwidth]{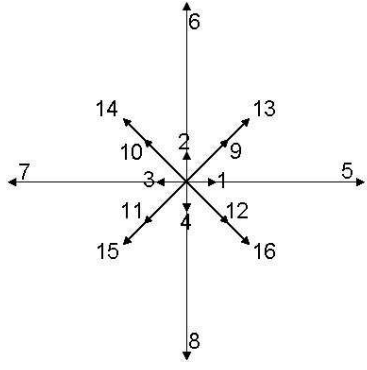}
\caption{Distribution of $\mathbf{v}_{i\protect\alpha }$ for the discrete
velocity model.}
\end{figure}

\subsubsection{Construction of transformation matrix $\mathbf{M}$}

The transformation matrix $\mathbf{M}$ is constructed according to the
irreducible representations of SO(2) group:
\begin{eqnarray*}
&&1\text{,} \\
&&\cos \theta \text{,}\sin \theta \text{,} \\
&&\sin ^{2}\theta +\cos ^{2}\theta \text{,}\cos 2\theta \text{,}\sin 2\theta\text{,} \\
&&\cos \theta (\sin ^{2}\theta +\cos ^{2}\theta )\text{,}\sin \theta (\sin^{2}\theta +\cos ^{2}\theta )\text{,}\cos 3\theta \text{,}\sin 3\theta \text{,} \\
&&(\sin ^{2}\theta +\cos ^{2}\theta )^{2}\text{,}\cos 4\theta \text{,}\cos 2\theta (\sin ^{2}\theta +\cos ^{2}\theta )\text{,}\\
&&\sin 2\theta (\sin^{2}\theta +\cos ^{2}\theta )\text{,} \\
&&\cos 3\theta (\sin ^{2}\theta +\cos ^{2}\theta )\text{,}\sin 3\theta (\sin^{2}\theta +\cos ^{2}\theta ) \text{,} \cdots
\end{eqnarray*}

Let $v_{ix}$ and $v_{iy}$ play the roles of $\cos \theta $ and $\sin \theta $%
, respectively. Then we define $m_{1i}=1$, $m_{2i}=v_{ix}$, $m_{3i}=v_{iy}$,
$m_{4i}=(v_{ix}^{2}+v_{iy}^{2})/2$, $m_{5i}=v_{ix}^{2}-v_{iy}^{2}$, $%
m_{6i}=v_{ix}v_{iy}$, $m_{7i}=v_{ix}(v_{ix}^{2}+v_{iy}^{2})/2$, $%
m_{8i}=v_{iy}(v_{ix}^{2}+v_{iy}^{2})/2$, $%
m_{9i}=v_{ix}(v_{ix}^{2}-3v_{iy}^{2})$, $%
m_{10i}=v_{iy}(3v_{ix}^{2}-v_{iy}^{2})$, $%
m_{11i}=(v_{ix}^{2}+v_{iy}^{2})^{2}/4$, $%
m_{12i}=v_{ix}^{4}-6v_{ix}^{2}v_{iy}^{2}+v_{iy}^{4}$, $%
m_{13i}=(v_{ix}^{2}+v_{iy}^{2})(v_{ix}^{2}-v_{iy}^{2})$, $%
m_{14i}=(v_{ix}^{2}+v_{iy}^{2})v_{ix}v_{iy}$, $%
m_{15i}=v_{ix}(v_{ix}^{2}-3v_{iy}^{2})(v_{ix}^{2}+v_{iy}^{2})$, $%
m_{16i}=v_{iy}(3v_{ix}^{2}-v_{iy}^{2})(v_{ix}^{2}+v_{iy}^{2})$, where $%
i=1,\cdots ,16$.

For two-dimensional compressible models, we have four conserved moments,
density $\hat{f}_{1}=\rho =\sum f_{i}m_{1i}$, momenta $\hat{f}%
_{2}=j_{x}=\rho u_{x}=\sum f_{i}m_{2i}$ and $\hat{f}_{3}=j_{y}=\rho
u_{y}=\sum f_{i}m_{3i}$, and energy $\hat{f}_{4}=e=\sum f_{i}m_{4i}$. To be
consistent with the idiomatic expression of energy, in the definitions of $%
m_{4i}$, $m_{7i}$ and $m_{8i}$, a coefficient $1/2$ is used. Similarly, a
coefficient $1/4$ is used in the definition of $m_{11i}$. The components of
transformation matrix $\mathbf{M}$\ are shown in table I.
\begin{widetext}
\begin{center}
\begin{table}
\caption{Transformation matrix of MRT-LB for compressible fluids.}
\begin{tabular}{ccccccccccccccccc}
\hline\hline $i$ & $m_{1i}$ & $m_{2i}$ & $m_{3i}$ & $m_{4i}$ & $m_{5i}$ & $m_{6i}$ & $%
m_{7i}$ & $m_{8i}$ & $m_{9i}$ & $m_{10i}$ & $m_{11i}$ & $m_{12i}$ & $m_{13i}$
& $m_{14i}$ & $m_{15i}$ & $m_{16i}$ \\ \hline
$1$ & $1$ & $1$ & $0$ & $\frac{1}{2}$ & $1$ & $0$ & $\frac{1}{2}$ & $0$ & $1$
& $0$ & $\frac{1}{4}$ & $1$ & $1$ & $0$ & $1$ & $0$ \\
$2$ & $1$ & $0$ & $1$ & $\frac{1}{2}$ & $-1$ & $0$ & $0$ & $\frac{1}{2}$ & $0
$ & $-1$ & $\frac{1}{4}$ & $1$ & $-1$ & $0$ & $0$ & $-1$ \\
$3$ & $1$ & $-1$ & $0$ & $\frac{1}{2}$ & $1$ & $0$ & $-\frac{1}{2}$ & $0$ & $%
-1$ & $0$ & $\frac{1}{4}$ & $1$ & $1$ & $0$ & $-1$ & $0$ \\
$4$ & $1$ & $0$ & $-1$ & $\frac{1}{2}$ & $-1$ & $0$ & $0$ & $-\frac{1}{2}$ &
$0$ & $1$ & $\frac{1}{4}$ & $1$ & $-1$ & $0$ & $0$ & $1$ \\
$5$ & $1$ & $6$ & $0$ & $18$ & $36$ & $0$ & $108$ & $0$ & $216$ & $0$ & $324$
& $1296$ & $1296$ & $0$ & $7776$ & $0$ \\
$6$ & $1$ & $0$ & $6$ & $18$ & $-36$ & $0$ & $0$ & $108$ & $0$ & $-216$ & $%
324$ & $1296$ & $-1296$ & $0$ & $0$ & $-7776$ \\
$7$ & $1$ & $-6$ & $0$ & $18$ & $36$ & $0$ & $-108$ & $0$ & $-216$ & $0$ & $%
324$ & $1296$ & $1296$ & $0$ & $-7776$ & $0$ \\
$8$ & $1$ & $0$ & $-6$ & $18$ & $-36$ & $0$ & $0$ & $-108$ & $0$ & $216$ & $%
324$ & $1296$ & $-1296$ & $0$ & $0$ & $7776$ \\
$9$ & $1$ & $\sqrt{2}$ & $\sqrt{2}$ & $2$ & $0$ & $2$ & $2\sqrt{2}$ & $2%
\sqrt{2}$ & $-4\sqrt{2}$ & $4\sqrt{2}$ & $4$ & $-16$ & $0$ & $8$ & $-16\sqrt{%
2}$ & $16\sqrt{2}$ \\
$10$ & $1$ & $-\sqrt{2}$ & $\sqrt{2}$ & $2$ & $0$ & $-2$ & $-2\sqrt{2}$ & $2%
\sqrt{2}$ & $4\sqrt{2}$ & $4\sqrt{2}$ & $4$ & $-16$ & $0$ & $-8$ & $16\sqrt{2%
}$ & $16\sqrt{2}$ \\
$11$ & $1$ & $-\sqrt{2}$ & $-\sqrt{2}$ & $2$ & $0$ & $2$ & $-2\sqrt{2}$ & $-2%
\sqrt{2}$ & $4\sqrt{2}$ & $-4\sqrt{2}$ & $4$ & $-16$ & $0$ & $8$ & $16\sqrt{2%
}$ & $-16\sqrt{2}$ \\
$12$ & $1$ & $\sqrt{2}$ & $-\sqrt{2}$ & $2$ & $0$ & $-2$ & $2\sqrt{2}$ & $-2%
\sqrt{2}$ & $-4\sqrt{2}$ & $-4\sqrt{2}$ & $4$ & $-16$ & $0$ & $-8$ & $-16%
\sqrt{2}$ & $-16\sqrt{2}$ \\
$13$ & $1$ & $\frac{3}{\sqrt{2}}$ & $\frac{3}{\sqrt{2}}$ & $\frac{9}{2}$ & $0
$ & $\frac{9}{2}$ & $\frac{27}{2\sqrt{2}}$ & $\frac{27}{2\sqrt{2}}$ & $-%
\frac{27}{\sqrt{2}}$ & $\frac{27}{\sqrt{2}}$ & $\frac{81}{4}$ & $-81$ & $0$
& $\frac{81}{2}$ & $-\frac{243}{\sqrt{2}}$ & $\frac{243}{\sqrt{2}}$ \\
$14$ & $1$ & $-\frac{3}{\sqrt{2}}$ & $\frac{3}{\sqrt{2}}$ & $\frac{9}{2}$ & $%
0$ & $-\frac{9}{2}$ & $-\frac{27}{2\sqrt{2}}$ & $\frac{27}{2\sqrt{2}}$ & $%
\frac{27}{\sqrt{2}}$ & $\frac{27}{\sqrt{2}}$ & $\frac{81}{4}$ & $-81$ & $0$
& $-\frac{81}{2}$ & $\frac{243}{\sqrt{2}}$ & $\frac{243}{\sqrt{2}}$ \\
$15$ & $1$ & $-\frac{3}{\sqrt{2}}$ & $-\frac{3}{\sqrt{2}}$ & $\frac{9}{2}$ &
$0$ & $\frac{9}{2}$ & $-\frac{27}{2\sqrt{2}}$ & $-\frac{27}{2\sqrt{2}}$ & $%
\frac{27}{\sqrt{2}}$ & $-\frac{27}{\sqrt{2}}$ & $\frac{81}{4}$ & $-81$ & $0$
& $\frac{81}{2}$ & $\frac{243}{\sqrt{2}}$ & $-\frac{243}{\sqrt{2}}$ \\
$16$ & $1$ & $\frac{3}{\sqrt{2}}$ & $-\frac{3}{\sqrt{2}}$ & $\frac{9}{2}$ & $%
0$ & $-\frac{9}{2}$ & $\frac{27}{2\sqrt{2}}$ & $-\frac{27}{2\sqrt{2}}$ & $-%
\frac{27}{\sqrt{2}}$ & $-\frac{27}{\sqrt{2}}$ & $\frac{81}{4}$ & $-81$ & $0$
& $-\frac{81}{2}$ & $-\frac{243}{\sqrt{2}}$ & $-\frac{243}{\sqrt{2}}$ \\
\hline\hline
\end{tabular}
\end{table}
\end{center}
\end{widetext}

\subsubsection{Determination of $\hat{f}_{i}^{eq}$}

The second group components of $\hat{f}_{i}^{eq}$ are chosen in such a way
that in the continuum limit the MRT LB model recovers the Navier-Stokes
equations. To that end, we perform the Chapman-Enskog expansion on the two
sides of Eq.\eqref{1}. We use the following multiscale expansions:
\begin{subequations}
\begin{equation}
f_{i}=f_{i}^{(0)}+f_{i}^{(1)}+f_{i}^{(2)}\text{,}  \label{6a}
\end{equation}%
\begin{equation}
\frac{\partial }{\partial t}=\frac{\partial }{\partial t_{1}}+\frac{\partial
}{\partial t_{2}}\text{,}  \label{6b}
\end{equation}%
\begin{equation}
\frac{\partial }{\partial x}=\frac{\partial }{\partial x_{1}}\text{,}
\label{6c}
\end{equation}%
where $f_{i}^{(0)}$\ is the zeroth order, $f_{i}^{(1)}$,\ $\frac{\partial }{%
\partial t_{1}}$and\ $\frac{\partial }{\partial x_{1}}$\ are the first
order, $f_{i}^{(2)}$\ and $\frac{\partial }{\partial t_{2}}$ are the second
order terms of the Knudsen number $\epsilon $. Equating the coefficients of
the zeroth, the first, and the second order terms in $\epsilon $ gives
\end{subequations}
\begin{subequations}
\begin{equation}
f_{i}^{(0)}=f_{i}^{eq}\text{,}  \label{7a}
\end{equation}%
\begin{equation}
(\frac{\partial }{\partial t_{1}}+v_{i\alpha }\frac{\partial }{\partial
x_{1\alpha }})f_{i}^{(0)}=-\mathbf{S}_{il}f_{l}^{(1)}\text{,}  \label{7b}
\end{equation}%
\begin{equation}
\frac{\partial }{\partial t_{2}}f_{i}^{(0)}+(\frac{\partial }{\partial t_{1}}%
+v_{i\alpha }\frac{\partial }{\partial x_{1\alpha }})f_{i}^{(1)}=-\mathbf{S}%
_{il}f_{l}^{(2)}\text{.}  \label{7c}
\end{equation}%
In the moment space they are
\end{subequations}
\begin{subequations}
\begin{equation}
\hat{f}_{i}^{(0)}=\hat{f}_{i}^{eq}\text{,}  \label{8a}
\end{equation}%
\begin{equation}
(\frac{\partial }{\partial t_{1}}+\hat{\mathbf{E}}_{\alpha }\frac{\partial }{%
\partial x_{1\alpha }})\hat{f}_{i}^{(0)}=-\hat{\mathbf{S}}_{il}\hat{f}%
_{l}^{(1)}\text{,}  \label{8b}
\end{equation}%
\begin{equation}
\frac{\partial }{\partial t_{2}}\hat{f}_{i}^{(0)}+(\frac{\partial }{\partial
t_{1}}+\hat{\mathbf{E}}_{\alpha }\frac{\partial }{\partial x_{1\alpha }})%
\hat{f}_{i}^{(1)}=-\hat{\mathbf{S}}_{il}\hat{f}_{l}^{(2)}\text{,}  \label{8c}
\end{equation}%
where $\hat{\mathbf{E}}_{\alpha }=\mathbf{M}(v_{i\alpha }\mathbf{I})\mathbf{M%
}^{-1}$.

The equilibria of the moments in the moment space read $\hat{\mathbf{f}}%
^{eq}=(\rho ,j_{x},j_{y},e,\hat{f}_{5}^{eq},\hat{f}_{6}^{eq},\cdots ,\hat{f}%
_{16}^{eq})^{T}$. The first and second order deviations from equilibria are
defined as : $\hat{\mathbf{f}}^{(1)}=(0,0,0,0,\hat{f}_{5}^{(1)},\hat{f}%
_{6}^{(1)},\cdots ,\hat{f}_{16}^{(1)})^{T}$ and $\hat{\mathbf{f}}%
^{(2)}=(0,0,0,0,\hat{f}_{5}^{(2)},\hat{f}_{6}^{(2)},\cdots ,\hat{f}%
_{16}^{(2)})^{T}$, respectively. Via some algebraic treatments, we obtain
\end{subequations}
\begin{subequations}
\begin{equation}
\frac{\partial \rho }{\partial t}+\frac{\partial j_{x}}{\partial x}+\frac{%
\partial j_{y}}{\partial y}=0\text{,}  \label{11a}
\end{equation}%
\begin{equation}
\frac{\partial j_{x}}{\partial t}+\frac{\partial }{\partial x}(e+\frac{1}{2}%
\hat{f}_{5}^{eq})+\frac{\partial }{\partial y}\hat{f}_{6}^{eq}=-\frac{1}{2}%
\frac{\partial }{\partial x}\hat{f}_{5}^{(1)}-\frac{\partial }{\partial y}%
\hat{f}_{6}^{(1)}\text{,}  \label{11b}
\end{equation}%
\begin{equation}
\frac{\partial j_{y}}{\partial t}+\frac{\partial }{\partial x}\hat{f}%
_{6}^{eq}+\frac{\partial }{\partial y}(e-\frac{1}{2}\hat{f}_{5}^{eq})=-\frac{%
\partial }{\partial x}\hat{f}_{6}^{(1)}+\frac{1}{2}\frac{\partial }{\partial
y}\hat{f}_{5}^{(1)}\text{,}  \label{11c}
\end{equation}%
\begin{equation}
\frac{\partial e}{\partial t}+\frac{\partial }{\partial x}\hat{f}_{7}^{eq}+%
\frac{\partial }{\partial y}\hat{f}_{8}^{eq}=-\frac{\partial }{\partial x}%
\hat{f}_{7}^{(1)}-\frac{\partial }{\partial y}\hat{f}_{8}^{(1)}\text{.}
\label{11d}
\end{equation}
\end{subequations}
If choose $\hat{f}_{5}^{eq}=(j_{x}^{2}-j_{y}^{2})/\rho $, $\hat{f}%
_{6}^{eq}=j_{x}j_{y}/\rho $, $\hat{f}_{7}^{eq}=(e+P)j_{x}/\rho $, $\hat{f}%
_{8}^{eq}=(e+P)j_{y}/\rho $, $\hat{f}_{9}^{eq}=(j_{x}^{2}-3j_{y}^{2})j_{x}/%
\rho ^{2}$, $\hat{f}_{10}^{eq}=(3j_{x}^{2}-j_{y}^{2})j_{y}/\rho ^{2}$, $\hat{%
f}_{11}^{eq}=2e^{2}/\rho -(j_{x}^{2}+j_{y}^{2})^{2}/4\rho ^{3}$, $\hat{f}%
_{13}^{eq}=(6\rho e-2j_{x}^{2}-2j_{y}^{2})(j_{x}^{2}-j_{y}^{2})/\rho ^{3}$, $%
\hat{f}_{14}^{eq}=(6\rho e-2j_{x}^{2}-2j_{y}^{2})j_{x}j_{y}/\rho ^{3}$, $%
\hat{f}_{12}^{eq}=\hat{f}_{15}^{eq}=\hat{f}_{16}^{eq}=0$, the MRT LB model
recovers the following Navier-Stokes equations:

\begin{widetext}
\begin{subequations}
\begin{equation}
\frac{\partial \rho }{\partial t}+\frac{\partial j_{x}}{\partial x}+\frac{%
\partial j_{y}}{\partial y}=0\text{,}  \label{15a}
\end{equation}%
\begin{equation}
\frac{\partial j_{x}}{\partial t}+\frac{\partial }{\partial x}\left(
j_{x}^{2}/\rho \right) +\frac{\partial }{\partial y}\left( j_{x}j_{y}/\rho
\right) =-\frac{\partial P}{\partial x}+\frac{\partial }{\partial x}[\mu
_{s}(\frac{\partial u_{x}}{\partial x}-\frac{\partial u_{y}}{\partial y})]+%
\frac{\partial }{\partial y}[\mu _{v}(\frac{\partial u_{y}}{\partial x}+%
\frac{\partial u_{x}}{\partial y})]\text{,}  \label{15b}
\end{equation}%
\begin{equation}
\frac{\partial j_{y}}{\partial t}+\frac{\partial }{\partial x}\left(
j_{x}j_{y}/\rho \right) +\frac{\partial }{\partial y}\left( j_{y}^{2}/\rho
\right) =-\frac{\partial P}{\partial y}+\frac{\partial }{\partial x}[\mu
_{v}(\frac{\partial u_{y}}{\partial x}+\frac{\partial u_{x}}{\partial y})]-%
\frac{\partial }{\partial y}[\mu _{s}(\frac{\partial u_{x}}{\partial x}-%
\frac{\partial u_{y}}{\partial y})]\text{,}  \label{15c}
\end{equation}%
\begin{eqnarray}
&&\frac{\partial e}{\partial t}+\frac{\partial }{\partial x}[(e+P)j_{x}/\rho
]+\frac{\partial }{\partial y}[(e+P)j_{y}/\rho ]  \notag \\
&=&\frac{\partial }{\partial x}[\lambda _{1}\frac{\partial T}{\partial x}+%
\frac{\lambda _{1}}{2}(u_{y}\frac{\partial u_{y}}{\partial x}+u_{x}\frac{%
\partial u_{x}}{\partial x}-u_{x}\frac{\partial u_{y}}{\partial y}+u_{y}%
\frac{\partial u_{x}}{\partial y})]  \notag \\
&&+\frac{\partial }{\partial y}[\lambda _{2}\frac{\partial T}{\partial y}+%
\frac{\lambda _{2}}{2}(u_{x}\frac{\partial u_{x}}{\partial y}-u_{y}\frac{%
\partial u_{x}}{\partial x}+u_{x}\frac{\partial u_{y}}{\partial x}+u_{y}%
\frac{\partial u_{y}}{\partial y})]\text{,}  \label{15d}
\end{eqnarray}
\end{subequations}
where $\mu _{s}=$ $\rho RT/s_{5}$, $\mu _{v}=$ $\rho RT/s_{6}$, $\lambda
_{1}=2\rho RT/s_{7}$, $\lambda _{2}=2\rho RT/s_{8}$. It is noted that the
definitions of $\hat{f}_{12}^{eq}$,\ $\hat{f}_{15}^{eq}$,\ $\hat{f}%
_{16}^{eq} $ have no effect on the recovered macroscopic equations. When $%
\mu _{s}=$\ $\mu _{v}=\mu $,\ $\lambda _{1}=\lambda _{2}=\lambda $, the
above Navier-Stokes equations reduce to
\begin{subequations}
\begin{equation}
\frac{\partial \rho }{\partial t}+\frac{\partial j_{\alpha }}{\partial
x_{\alpha }}=0\text{,}  \label{16a}
\end{equation}%
\begin{equation}
\frac{\partial j_{\alpha }}{\partial t}+\frac{\partial \left( j_{\alpha
}j_{\beta }/\rho \right) }{\partial x_{\beta }}=-\frac{\partial P}{\partial
x_{\alpha }}+\frac{\partial }{\partial x_{\beta }}[\mu (\frac{\partial
u_{\alpha }}{\partial x_{\beta }}+\frac{\partial u_{\beta }}{\partial
x_{\alpha }}-\frac{\partial u_{\chi }}{\partial x_{\chi }}\delta _{\alpha
\beta })]\text{,}  \label{16b}
\end{equation}%
\begin{equation}
\frac{\partial e}{\partial t}+\frac{\partial }{\partial x_{\alpha }}%
[(e+P)j_{\alpha }/\rho ]=\frac{\partial }{\partial x_{\alpha }}[\lambda
\frac{\partial T}{\partial x_{\alpha }}+\frac{\lambda }{2}u_{\beta }(\frac{%
\partial u_{\alpha }}{\partial x_{\beta }}+\frac{\partial u_{\beta }}{%
\partial x_{\alpha }}-\frac{\partial u_{\chi }}{\partial x_{\chi }}\delta
_{\alpha \beta })]\text{.}  \label{16c}
\end{equation}
\end{subequations}
\end{widetext}
%

In Fig.6(a) we show an example of stability comparison for the new MRT model
and its SRT version. The abscissa is for $kdx$, and the vertical axis is for
$|\omega |_{max}$\ which is the largest eigenvalue of coefficient matrix $%
G_{ij}$. The macroscopic values in stability analysis are chosen as follows:
$(\rho ,u_{x},u_{y},T)$ = $(2.0,10.0,0.0,2.0)$. The relaxation time in SRT
is $\tau =10^{-5}$, while the collision parameters in MRT are $s_{5}=6500$, $%
s_{7}=s_{8}=9\times 10^{4}$, $s_{9}=8\times 10^{4}$, $s_{13}=7\times 10^{4}$%
, $s_{14}=8\times 10^{3}$, $s_{15}=2.5\times 10^{4}$, the others are $10^{5}$%
. In this case, the MRT scheme is stable, while the SRT version is not. It
is clear that, by choosing appropriate collision parameters, the stability
of MRT can be much better than the SRT.

\begin{figure}[tbp]
\center\includegraphics*[bbllx=7pt,bblly=107pt,bburx=575pt,bbury=403pt,angle=0,width=0.78\textwidth]
{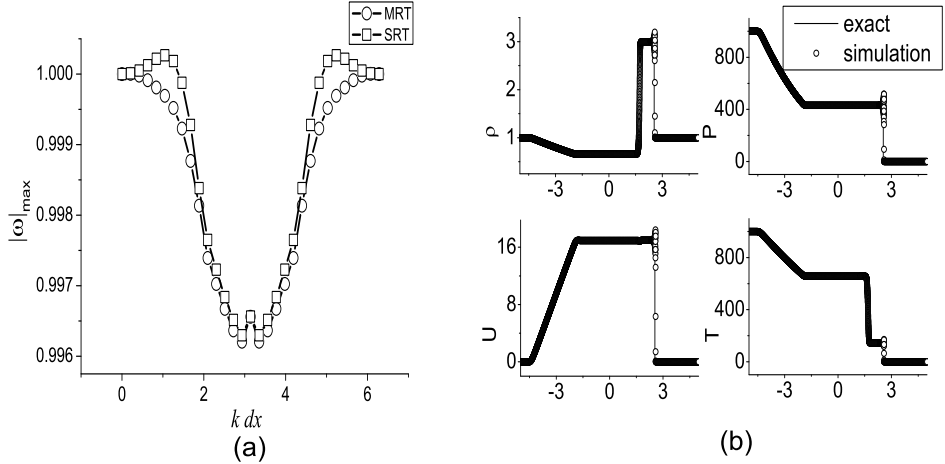}
\caption{(a) Stability comparison for the new MRT model and its SRT version.
(b) The MRT simulation results and exact solutions for the Colella explosion
wave at time $t=0.1$.}
\end{figure}

Figure 6(b) shows the comparison of MRT LB results and exact ones for the
well-known Colella explosion wave problem. For the problem, the initial
condition is
\begin{equation}
\left\{
\begin{array}{cc}
(\rho ,u_{x},u_{y},T)|_{L}=(1.0,0.0,0.0,1000.0)\text{,} & x\leq 0\text{.} \\
(\rho ,u_{x},u_{y},T)|_{R}=(1.0,0.0,0.0,0.01)\text{,} & x>0\text{.}%
\end{array}%
\right.
\end{equation}%
This is a strong temperature discontinuity problem that can be used to study
the robustness and precision of numerical methods. Figure 6(b) gives density,
pressure, velocity and temperature results at $t=0.1$. Symbols are for
simulation results. Here, the parameters are $s_{7}=s_{8}=5\times 10^{4}$, $%
s_{11}=s_{13}=5\times 10^{5}$, other values of $s$\ still adopt $10^{5}$.
The success of the simulation shows that the MRT model is applicable to
simulate strong temperature discontinuity problem, and confirms the
robustness and precision of the model.

Two points should be commented here. The first is that the better stability
is not the only or most important advantage of MRT over SRT. From the view
of physical modeling, the SRT is only a special case of the MRT. The second
is that the above MRT LB model works well for shocked compressible fluids
where the shocking procedure is much faster than the transportation
processes. To work also well for more general cases, the collision operators
of the moments related to the energy flux should be modified as below \cite%
{MRT2011TAML},
\begin{widetext}
\begin{equation}
\hat{\mathbf{S}}_{77}(\hat{f}_{7}-\hat{f}_{7}^{eq})\Rightarrow
\hat{\mathbf{S%
}}_{77}(\hat{f}_{7}-\hat{f}_{7}^{eq})+(s_{7}/s_{5}-1)\rho Tu_{x}(\frac{%
\partial u_{x}}{\partial x}-\frac{\partial u_{y}}{\partial y}%
)+(s_{7}/s_{6}-1)\rho Tu_{y}(\frac{\partial u_{y}}{\partial x}+\frac{%
\partial u_{x}}{\partial y})\text{,}  \label{modify1}
\end{equation}%
\begin{equation}
\hat{\mathbf{S}}_{88}(\hat{f}_{8}-\hat{f}_{8}^{eq})\Rightarrow
\hat{\mathbf{S%
}}_{88}(\hat{f}_{8}-\hat{f}_{8}^{eq})+(s_{8}/s_{6}-1)\rho Tu_{x}(\frac{%
\partial u_{y}}{\partial x}+\frac{\partial u_{x}}{\partial y}%
)-(s_{8}/s_{5}-1)\rho Tu_{y}(\frac{\partial u_{x}}{\partial x}-\frac{%
\partial u_{y}}{\partial y})\text{.}  \label{modify2}
\end{equation}

After the modification the coefficients of viscosity in energy equation %
\eqref{15d} are consistent with those in momentum equations \eqref{15b}-%
\eqref{15c}.

\subsection{MRT model based on moment relations}

In the original KT model, besides Eqs. \eqref{e4}-\eqref{e8}, the local
equilibrium distribution function $f_{i}^{eq}$ is required to satisfy the
following two additional moment relations:
\begin{subequations}
\begin{equation}
\rho \left[ RT\left( u_{\alpha }\delta _{\beta \chi }+u_{\beta }\delta
_{\alpha \chi }+u_{\chi }\delta _{\alpha \beta }\right) +u_{\alpha }u_{\beta
}u_{\chi }\right] =\sum f_{i}^{eq}v_{i\alpha }v_{i\beta }v_{i\chi },
\label{4f}
\end{equation}%
\begin{equation}
\rho \left\{ \left( b+2\right) R^{2}T^{2}\delta _{\alpha \beta }+\left[
\left( b+4\right) u_{\alpha }u_{\beta }+u_{\chi }^{2}\delta _{\alpha \beta }%
\right] RT+u_{\chi }^{2}u_{\alpha }u_{\beta }\right\} =\sum f_{i}^{eq}\left(
v_{i\chi }^{2}+\eta _{i}^{2}\right) v_{i\alpha }v_{i\beta }  \label{4g}
\end{equation}%
The local equilibrium distribution function \ $f_{i}^{eq}$\ is calculated
via the following polynomial:
\end{subequations}
\begin{eqnarray}
f_{i}^{eq} &=&\rho [ a_{0i}+a_{1i}T+a_{2i}T^{2}+\left( a_{3i}+a_{4i}T\right)
u_{\alpha }^{2}+a_{5i}u_{\alpha }^{2}u_{\beta }^{2}  \notag \\
&&+\left( b_{0i}+b_{1i}T+b_{2i}u_{\alpha }^{2}\right) u_{\beta }v_{i\beta
}+\left( d_{0i}+d_{1i}T+d_{2i}u_{\alpha }^{2}\right) u_{\beta }v_{i\beta
}u_{\chi }v_{i\chi }  \notag \\
&&+e_{i}u_{\alpha }v_{i\alpha }u_{\beta }v_{i\beta }u_{\chi }v_{i\chi }]%
\text{,}  \label{eq}
\end{eqnarray}%
which is of the flow velocity up to the third order. The coefficients $%
a_{0i} $ , $\ldots $, $e_{i}$\ ( $i=1,\ldots ,16$ ) in the distribution
function $f_{i}^{eq}$\ are referred to the original publication \cite%
{Kataoka2004b}.

\subsubsection{Construction of transformation matrix $\mathbf{M}$}

In this MRT model the moments are chosen according to the seven required
moment relations \cite{MRT2010EPL,MRT2011CTP}. The RHS of the seven equations
indicate seven monomials: $1$, $v_{i\alpha }$, $v_{i\alpha }^{2}+\eta
_{i}^{2}$, $v_{i\alpha }v_{i\beta }$, $(v_{i\beta }^{2}+\eta
_{i}^{2})v_{i\alpha }$, $v_{i\alpha }v_{i\beta }v_{i\chi }$, $(v_{i\chi
}^{2}+\eta _{i}^{2})v_{i\alpha }v_{i\beta }$. Three possibilities arise from
the monomial $v_{i\alpha }v_{i\beta }$: (a) $\alpha =\beta =x$, $v_{i\alpha
}v_{i\beta }=v_{ix}^{2}$, (b) $\alpha =\beta =y$, $v_{i\alpha }v_{i\beta
}=v_{iy}^{2}$, (c) $\alpha =x$, $\beta =y$, $v_{i\alpha }v_{i\beta
}=v_{ix}v_{iy}$. \textquotedblleft (a)+(b)" gives $(v_{ix}^{2}+v_{iy}^{2})$,
\textquotedblleft (a)-(b)" gives $(v_{ix}^{2}-v_{iy}^{2})$. Through such a
simple combination of these monomials, we can compose the transformation
matrix $\mathbf{M}$ as below: $m_{1i}=1$, $m_{2i}=v_{ix}$, $m_{3i}=v_{iy}$, $%
m_{4i}=v_{ix}^{2}+v_{iy}^{2}+\eta _{i}^{2}$, $m_{5i}=v_{ix}^{2}+v_{iy}^{2}$,
$m_{6i}=v_{ix}^{2}-v_{iy}^{2}$, $m_{7i}=v_{ix}v_{iy}$, $%
m_{8i}=v_{ix}(v_{ix}^{2}+v_{iy}^{2}+\eta _{i}^{2})$, $%
m_{9i}=v_{iy}(v_{ix}^{2}+v_{iy}^{2}+\eta _{i}^{2})$, $%
m_{10i}=v_{ix}(v_{ix}^{2}+v_{iy}^{2})$, $%
m_{11i}=v_{iy}(v_{ix}^{2}+v_{iy}^{2})$, $%
m_{12i}=v_{ix}(v_{ix}^{2}-v_{iy}^{2})$, $%
m_{13i}=v_{iy}(v_{ix}^{2}-v_{iy}^{2})$, $%
m_{14i}=(v_{ix}^{2}+v_{iy}^{2})(v_{ix}^{2}+v_{iy}^{2}+\eta _{i}^{2})$, $%
m_{15i}=v_{ix}v_{iy}(v_{ix}^{2}+v_{iy}^{2}+\eta _{i}^{2})$, $%
m_{16i}=(v_{ix}^{2}-v_{iy}^{2})(v_{ix}^{2}+v_{iy}^{2}+\eta _{i}^{2})$, where
$i=1,\cdots ,16$. The components of transformation matrix $\mathbf{M}$\ are
shown in table II.

\begin{center}
\begin{table}[tbp]
\caption{Transformation matrix of MRT-LB model for compressible flows with
flexible specific-heat ratio.}%
\begin{tabular}{ccccccccccccccccc}
\hline\hline
$i$ & $m_{1i}$ & $m_{2i}$ & $m_{3i}$ & $m_{4i}$ & $m_{5i}$ & $m_{6i}$ & $%
m_{7i}$ & $m_{8i}$ & $m_{9i}$ & $m_{10i}$ & $m_{11i}$ & $m_{12i}$ & $m_{13i}$
& $m_{14i}$ & $m_{15i}$ & $m_{16i}$ \\ \hline
$1$ & $1$ & $1$ & $0$ & $\frac{29}{4}$ & $1$ & $1$ & $0$ & $\frac{29}{4}$ & $%
0$ & $1$ & $0$ & $1$ & $0$ & $\frac{29}{4}$ & $0$ & $\frac{29}{4}$ \\
$2$ & $1$ & $0$ & $1$ & $\frac{29}{4}$ & $1$ & $-1$ & $0$ & $0$ & $\frac{29}{%
4}$ & $0$ & $1$ & $0$ & $-1$ & $\frac{29}{4}$ & $0$ & $-\frac{29}{4}$ \\
$3$ & $1$ & $-1$ & $0$ & $\frac{29}{4}$ & $1$ & $1$ & $0$ & $-\frac{29}{4}$
& $0$ & $-1$ & $0$ & $-1$ & $0$ & $\frac{29}{4}$ & $0$ & $\frac{29}{4}$ \\
$4$ & $1$ & $0$ & $-1$ & $\frac{29}{4}$ & $1$ & $-1$ & $0$ & $0$ & $-\frac{29%
}{4}$ & $0$ & $-1$ & $1$ & $1$ & $\frac{29}{4}$ & $0$ & $-\frac{29}{4}$ \\
$5$ & $1$ & $6$ & $0$ & $36$ & $36$ & $36$ & $0$ & $216$ & $0$ & $216$ & $0$
& $216$ & $0$ & $1296$ & $0$ & $1296$ \\
$6$ & $1$ & $0$ & $6$ & $36$ & $36$ & $-36$ & $0$ & $0$ & $216$ & $0$ & $216$
& $0$ & $-216$ & $1296$ & $0$ & $-1296$ \\
$7$ & $1$ & $-6$ & $0$ & $36$ & $36$ & $36$ & $0$ & $-216$ & $0$ & $-216$ & $%
0$ & $216$ & $0$ & $1296$ & $0$ & $1296$ \\
$8$ & $1$ & $0$ & $-6$ & $36$ & $36$ & $-36$ & $0$ & $0$ & $-216$ & $0$ & $%
-216$ & $0$ & $216$ & $1296$ & $0$ & $-1296$ \\
$9$ & $1$ & $\sqrt{2}$ & $\sqrt{2}$ & $4$ & $4$ & $0$ & $2$ & $4\sqrt{2}$ & $%
4\sqrt{2}$ & $4\sqrt{2}$ & $4\sqrt{2}$ & $0$ & $0$ & $16$ & $8$ & $0$ \\
$10$ & $1$ & $-\sqrt{2}$ & $\sqrt{2}$ & $4$ & $4$ & $0$ & $-2$ & $-4\sqrt{2}$
& $4\sqrt{2}$ & $-4\sqrt{2}$ & $4\sqrt{2}$ & $0$ & $0$ & $16$ & $-8$ & $0$
\\
$11$ & $1$ & $-\sqrt{2}$ & $-\sqrt{2}$ & $4$ & $4$ & $0$ & $2$ & $-4\sqrt{2}$
& $-4\sqrt{2}$ & $-4\sqrt{2}$ & $-4\sqrt{2}$ & $0$ & $0$ & $16$ & $8$ & $0$
\\
$12$ & $1$ & $\sqrt{2}$ & $-\sqrt{2}$ & $4$ & $4$ & $0$ & $-2$ & $4\sqrt{2}$
& $-4\sqrt{2}$ & $4\sqrt{2}$ & $-4\sqrt{2}$ & $0$ & $0$ & $16$ & $-8$ & $0$
\\
$13$ & $1$ & $\frac{3}{\sqrt{2}}$ & $\frac{3}{\sqrt{2}}$ & $9$ & $9$ & $0$ &
$\frac{9}{2}$ & $\frac{27}{\sqrt{2}}$ & $\frac{27}{\sqrt{2}}$ & $\frac{27}{%
\sqrt{2}}$ & $\frac{27}{\sqrt{2}}$ & $0$ & $0$ & $81$ & $\frac{81}{2}$ & $0$
\\
$14$ & $1$ & $-\frac{3}{\sqrt{2}}$ & $\frac{3}{\sqrt{2}}$ & $9$ & $9$ & $0$
& $-\frac{9}{2}$ & $-\frac{27}{\sqrt{2}}$ & $\frac{27}{\sqrt{2}}$ & $-\frac{%
27}{\sqrt{2}}$ & $\frac{27}{\sqrt{2}}$ & $0$ & $0$ & $81$ & $-\frac{81}{2}$
& $0$ \\
$15$ & $1$ & $-\frac{3}{\sqrt{2}}$ & $-\frac{3}{\sqrt{2}}$ & $9$ & $9$ & $0$
& $\frac{9}{2}$ & $-\frac{27}{\sqrt{2}}$ & $-\frac{27}{\sqrt{2}}$ & $-\frac{%
27}{\sqrt{2}}$ & $-\frac{27}{\sqrt{2}}$ & $0$ & $0$ & $81$ & $\frac{81}{2}$
& $0$ \\
$16$ & $1$ & $\frac{3}{\sqrt{2}}$ & $-\frac{3}{\sqrt{2}}$ & $9$ & $9$ & $0$
& $-\frac{9}{2}$ & $\frac{27}{\sqrt{2}}$ & $-\frac{27}{\sqrt{2}}$ & $\frac{27%
}{\sqrt{2}}$ & $-\frac{27}{\sqrt{2}}$ & $0$ & $0$ & $81$ & $-\frac{81}{2}$ &
$0$ \\ \hline\hline
\end{tabular}%
\end{table}
\end{center}

It should be pointed out that, different from the other MRT models for
isothermal fluids, the transformation matrix $\mathbf{M}$ should not be
based upon a Gram-Schmidt orthogonalization procedure.

\subsubsection{Determination of $\hat{f}_{i}^{eq}$}

The procedure of determining $\hat{\mathbf{f}}^{eq}$ is similar to that for
the first MRT LB model in this paper. But the results are significantly
different. Our choice for this model is as below: $\hat{\mathbf{f}}%
^{eq}=(\rho ,j_{x},j_{y},e^{\prime },\hat{f}_{5}^{eq},\hat{f}%
_{6}^{eq},\cdots ,\hat{f}_{16}^{eq})^{T}$, where $\hat{f}%
_{5}^{eq}=2P+(j_{x}^{2}+j_{y}^{2})/\rho $, $\hat{f}%
_{6}^{eq}=(j_{x}^{2}-j_{y}^{2})/\rho $, $\hat{f}_{7}^{eq}=j_{x}j_{y}/\rho $,
$\hat{f}_{8}^{eq}=(e^{\prime }+2P)j_{x}/\rho $, $\hat{f}_{9}^{eq}=(e^{\prime
}+2P)j_{y}/\rho $, $\hat{f}_{10}^{eq}=(4P+j_{x}^{2}/\rho +j_{y}^{2}/\rho
)j_{x}/\rho $, $\hat{f}_{11}^{eq}=(4P+j_{x}^{2}/\rho +j_{y}^{2}/\rho
)j_{y}/\rho $, $\hat{f}_{12}^{eq}=(2P+j_{x}^{2}/\rho -j_{y}^{2}/\rho
)j_{x}/\rho $, $\hat{f}_{13}^{eq}=(-2P+j_{x}^{2}/\rho -j_{y}^{2}/\rho
)j_{y}/\rho $, $\hat{f}_{14}^{eq}=2(b+2)\rho
R^{2}T^{2}+(6+b)RT(j_{x}^{2}+j_{y}^{2})/\rho +(j_{x}^{2}+j_{y}^{2})^{2}/\rho
^{3}$, $\hat{f}_{15}^{eq}=[(b+4)P+(j_{x}^{2}+j_{y}^{2})/\rho
]j_{x}j_{y}/\rho ^{2}$, $\hat{f}_{16}^{eq}=[(b+4)P+(j_{x}^{2}+j_{y}^{2})/%
\rho ](j_{x}^{2}-j_{y}^{2})/\rho ^{2}$, where $P=\rho RT$, and $e^{\prime
}=b\rho RT+j_{\alpha }^{2}/\rho $ is the twice of total energy $e$. The
recovered Navier-Stokes equations are as follows:
\begin{subequations}
\begin{equation}
\frac{\partial \rho }{\partial t}+\frac{\partial j_{x}}{\partial x}+\frac{%
\partial j_{y}}{\partial y}=0,  \label{12a}
\end{equation}%
\begin{eqnarray}
&&\frac{\partial j_{x}}{\partial t}+\frac{\partial }{\partial x}\left( \frac{%
j_{x}^{2}}{\rho }\right) +\frac{\partial }{\partial y}\left( \frac{j_{x}j_{y}%
}{\rho }\right)  \notag \\
&=&-\frac{\partial P}{\partial x}+\frac{\partial }{\partial y}[\frac{\rho RT%
}{s_{7}}(\frac{\partial u_{y}}{\partial x}+\frac{\partial u_{x}}{\partial y}%
)]  \notag \\
&&+\frac{\partial }{\partial x}[\frac{\rho RT}{s_{5}}(1-\frac{2}{b})(\frac{%
\partial u_{x}}{\partial x}+\frac{\partial u_{y}}{\partial y})+\frac{\rho RT%
}{s_{6}}(\frac{\partial u_{x}}{\partial x}-\frac{\partial u_{y}}{\partial y}%
)]\text{,}  \label{12b}
\end{eqnarray}%
\begin{eqnarray}
&&\frac{\partial j_{y}}{\partial t}+\frac{\partial }{\partial x}\left( \frac{%
j_{x}j_{y}}{\rho }\right) +\frac{\partial }{\partial y}\left( \frac{j_{y}^{2}%
}{\rho }\right)  \notag \\
&=&-\frac{\partial P}{\partial y}+\frac{\partial }{\partial x}[\frac{\rho RT%
}{s_{7}}(\frac{\partial u_{y}}{\partial x}+\frac{\partial u_{x}}{\partial y}%
)]  \notag \\
&&+\frac{\partial }{\partial y}[\frac{\rho RT}{s_{5}}(1-\frac{2}{b})(\frac{%
\partial u_{x}}{\partial x}+\frac{\partial u_{y}}{\partial y})-\frac{\rho RT%
}{s_{6}}(\frac{\partial u_{x}}{\partial x}-\frac{\partial u_{y}}{\partial y}%
)]\text{,}  \label{12c}
\end{eqnarray}%
\begin{eqnarray}
&&\frac{\partial e}{\partial t}+\frac{\partial }{\partial x}[(e+P)j_{x}/\rho
]+\frac{\partial }{\partial y}[(e+P)j_{y}/\rho ]  \notag \\
&=&\frac{\partial }{\partial x}\{\frac{\rho RT}{s_{8}}[(\frac{b}{2}+1)R\frac{%
\partial T}{\partial x}+(2\frac{\partial u_{x}}{\partial x}-\frac{2}{b}\frac{%
\partial u_{x}}{\partial x}-\frac{2}{b}\frac{\partial u_{y}}{\partial y}%
)u_{x}+(\frac{\partial u_{y}}{\partial x}+\frac{\partial u_{x}}{\partial y}%
)u_{y}]\}  \notag \\
&&+\frac{\partial }{\partial y}\{\frac{\rho RT}{s_{9}}[(\frac{b}{2}+1)R\frac{%
\partial T}{\partial y}+(2\frac{\partial u_{y}}{\partial y}-\frac{2}{b}\frac{%
\partial u_{x}}{\partial x}-\frac{2}{b}\frac{\partial u_{y}}{\partial y}%
)u_{y}+(\frac{\partial u_{y}}{\partial x}+\frac{\partial u_{x}}{\partial y}%
)u_{x}]\}\text{.}  \label{12d}
\end{eqnarray}%
When $s_{5}=s_{6}=s_{7}=s_{8}=s_{9}$, the above Navier-Stokes equations
reduce to
\end{subequations}
\begin{subequations}
\begin{equation}
\frac{\partial \rho }{\partial t}+\frac{\partial j_{\alpha }}{\partial
x_{\alpha }}=0\text{,}  \label{16a1}
\end{equation}%
\begin{equation}
\frac{\partial j_{\alpha }}{\partial t}+\frac{\partial \left( j_{\alpha
}j_{\beta }/\rho \right) }{\partial x_{\beta }}=-\frac{\partial P}{\partial
x_{\alpha }}-\frac{\partial }{\partial x_{\beta }}P_{\alpha \beta
}^{^{\prime }}\text{,}  \label{16b1}
\end{equation}%
\begin{equation}
\frac{\partial e}{\partial t}+\frac{\partial }{\partial x_{\alpha }}\left[
(e+P)u_{\alpha }\right] =\frac{\partial }{\partial x_{\beta }}\left( (\frac{b%
}{2}+1)\mu R\frac{\partial T}{\partial x_{\beta }}-P_{\alpha \beta
}^{^{\prime }}u_{\alpha }\right) \text{,}  \label{16c1}
\end{equation}%
where
\end{subequations}
\begin{equation*}
\mu =\frac{\rho RT}{s}\text{,}\mu _{B}=(2/3-2/b)\frac{\rho RT}{s}\text{,}
\end{equation*}%
\begin{equation*}
P_{\alpha \beta }^{^{\prime }}=-\mu \left( \frac{\partial u_{\alpha }}{%
\partial x_{\beta }}+\frac{\partial u_{\beta }}{\partial x_{\alpha }}-\frac{2%
}{3}\frac{\partial u_{\chi }}{\partial x_{\chi }}\delta _{\alpha \beta
}\right) -\mu _{B}\frac{\partial u_{\chi }}{\partial x_{\chi }}\delta
_{\alpha \beta }\text{,}\left( \alpha ,\beta ,\gamma =x,y\right) \text{.}
\end{equation*}

Similar to the case of the first MRT model, the second one works also well
for shocked compressible fluids. For more general cases, similar
modifications to the collision operators of the moments related to the
energy flux should be made \cite{MRT2011TAML}.
\end{widetext}

\section{Simulations on hydrodynamic instabilities}

Hydrodynamic instabilities are ubiquitous in natural and industrial
processes. The Rayleigh-Taylor(RT) instability, Richtmyer-Meshkov(RM)
instability and Kelvin-Helmholtz (KH) instability are highly concerned in
weapon physics and inertial confinement fusion. For example, during the
spherical implosion procedure, the high pressure applied at the outside of
the shell drives a very strong shock wave towards the centre of the device.
This shock wave first accelerates the interface to a high velocity. Towards
the end of the implosion the interface is decelerated by a combination of
shock waves reflected from the center of the device and continuous
deceleration due to build up of high pressure in the thermonuclear material.
Such a very complicated acceleration/deceleration behavior results in two
processes, RT instability and RM instability. Since the implosion is
generally not perfectly symmetrical, the shear at the interface induces the
third process, KH instability. Hydrodynamic instabilities in such procedure
influence significantly the implosion physics and weapon performance. In
this section, we summarize our recent attempts on LB simulations on KH \cite%
{LB2KHI2011} and RM instabilities \cite{MRT2010EPL,MRT2011CTP}. When studying
the RM instability, the system must be compressible. In the case of KH
instability, the system can be compressible or nearly incompressible. As a
first step, we attempted the case with nearly incompressible fluids.

\subsection{Richtmyer-Meshkov instability}

The RM instability arises when a shock wave interacts with an interface
separating two different fluids. It combines various compressible phenomena,
such as shock interaction and refraction, with hydrodynamic instability,
including nonlinear growth and subsequent transition to turbulence, across a
wide range of Mach numbers. In inertial confinement fusion, the RM
instability causes mixing between the capsule material and the fuel within,
limiting final compression and thus the ability to achieve energy break-even
or production. The RM instability problems in the plane occur when a shock
wave travels from a light medium to a heavy one or when the shock wave
travels from a heavy medium to a light one.

\subsubsection{Shock wave from light to heavy media}

A practical example for this case is that the shock wave travels from air to
$SF_{6}$. For such a case, in our LB simulations we set the following
initial physical field,
\begin{equation*}
\left\{
\begin{array}{cc}
(\rho ,u_{x},u_{y},p)_{l}=(1.34161,0.361538,0,1.51332)\text{,} &  \\
(\rho ,u_{x},u_{y},p)_{m}=(1,0,0,1)\text{,} &  \\
(\rho ,u_{x},u_{y},p)_{r}=(5.04,0,0,1)\text{,} &
\end{array}%
\right.
\end{equation*}%
where the subscripts $l$, $m$, $r$ indicate the left, middle, right regions
of the whole domain. Such an initial configuration can be explained as
below: the interface of the middle and right regions separates the light and
heavy media; the interface of the left and middle regions is the shock
front; the shocked light medium is in the left and the pre-shocked is in the
middle regions. Initially, the two media have the same pressure and
different densities and temperatures. The corresponding Mach number of the
shock wave traveling from left is $1.2$. The shock wave will hit the
interface with an initial sinusoidal perturbation. The initial sinusoidal
perturbation at the interface reads $x=0.25\times N_{x}\times dx+0.008\times
\cos (20\pi y)$, where the cycle in $y$-direction of initial perturbation is
$0.1$, the amplitude is $0.008$, $N_{x}$ is grid number, and $dx$ is grid
size. The following boundary conditions are imposed: (i) inflow at the left
side; (ii) outflow at the right side, and (3) periodic in the $y$%
-directions. $\gamma =1.4$ in the whole domain.

Since the Mach number is $1.2$, the compressibility effects in this case is
not negligible. Figure 7 shows the density and pressure contours at four
different times, $t=0$, $0.06$, $0.3$ and $1.15$. When the shock wave passes
the interface from the left, a reflected shock wave to the left and a
transmission wave to the right are generated (clearly seen in pressure field
at $t=0.06$). The transmission wave has a certain curvature at this time.
Due to the compression, the interface produces a small deformation, and the
perturbation amplitude reduces slightly. At $t=0.3$, the reflected shock
wave has been out of the computational domain, and the transmission wave
becomes flat, which is consistent with the theoretical analysis of  \cite%
{Velikovich}. The perturbation amplitude begins to increase under the
pressure gradient, producing the bubble and spike structures. The
misalignment of pressure and density gradients causes a deposition of
vorticity at the top of spike structure, and a mushroom shape is formed
eventually. Fig.8 shows the changes of perturbation amplitude and growth
rate with time. The amplitude is defined as half of the maximum distance
between the crest and trough. From Fig.8 one can clearly find the initial
decrease of perturbation amplitude. During this initial period, the growth
rate is negative.

\begin{figure}[tbp]
\center\includegraphics*[ width=0.68\textwidth]{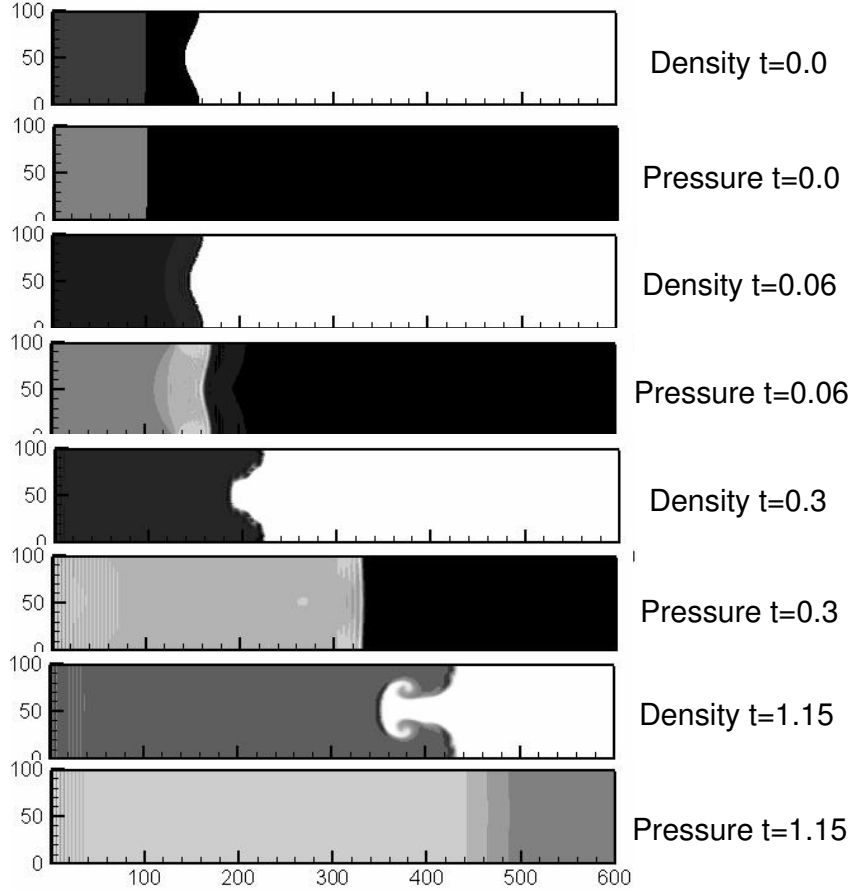}
\caption{Snapshots of RM instability (from light to heavy medium): density
and pressure contours at $t=0$, $t=0.06$, $t=0.3$, $t=1.15$, respectively.
From deep to light color, the level corresponds to the increase of values.}
\end{figure}

\begin{figure}[tbp]
\includegraphics*[ width=0.78\textwidth]{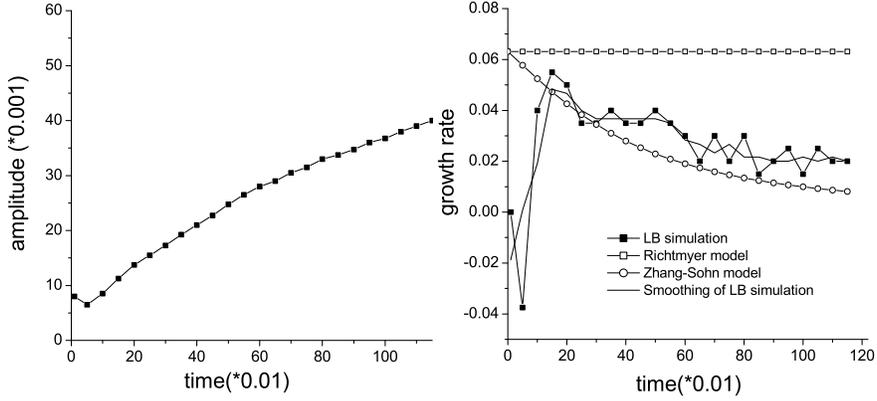}
\caption{Amplitude and growth rate changes with time (from light to heavy
medium).}
\end{figure}

Now we go to some theories to explain and validate the simulation results.
Richtmyer \cite{Richtmyer1960} proposed an impulsive model in the case of a
reflected shock wave via modifying the linear theory of Taylor for
Rayleigh-Taylor instability. According to the impulse model, the growth rate
reads,
\begin{equation*}
\frac{da}{dt}=k\Delta uA_{1}a_{1}\text{,}\qquad a_{1}=a_{0}(1-\frac{\Delta u%
}{D})
\end{equation*}%
where $k=2\pi /\lambda $ is the wave number, $\Delta u$ is the velocity
change across the interface, $A_{1}$ is the post-shock Atwood number, $a_{1}$
represents the post-shock amplitude, $a_{0}$ is the initial amplitude, $D$
denotes the incident shock speed. $Cmpr=1-\Delta u/D$ is compression ratio.
According to the initial conditions, the solution is $Cmpr=0.84$, $%
da/dt=0.063$. In the experiments of Meshkov \cite{Meshkov69} and Benjamin \cite%
{Benjamin92}, the measured growth rates are only about one half of that
predicted by the impulsive model. Zhang and Sohn \cite{Zhang} developed a
model for the growth of RM unstable interface from early to late times in
the case of light-heavy transition. The amplitude growth reads
\begin{equation*}
\frac{da}{dt}=\frac{v_{0}}{1+k^{2}v_{0}a_{1}t+\max
[0,(ka_{1})^{2}-(A_{1})^{2}+0.5](kv_{0}t)^{2}}
\end{equation*}%
where $v_{0}=k\Delta uA_{1}a_{1}$. As shown in Fig.8, the LB result for
growth rate qualitatively agrees well with that of Zhang-Sohn model. The
amplitude reaches the minimum value $0.0065$ at time $t=0.05$, so the
compression ratio obtained in simulation is $Cmpr=0.0065/0.008=0.81$. By the
least squares fitting, the growth rate of amplitude $0.03$ is obtained,
which is about one half of the growth rate predicted by the impulsive model
and consequently is in good agreement with the experimental result. In the nonlinear
stage, the simulation results agree qualitatively well with the perturbation
model proposed by Zhang and Sohn.

\subsubsection{Shock wave from heavy to light media}

A practical example is that the shock wave travels from air to \emph{He}. To better
understand such a case, in our LB simulation, we set a planar shock wave
with the Mach number $2.5$ impinging on a sinusoidal perturbation $%
x=0.1+0.008\times \cos (20\pi y)$, where the cycle and amplitude of initial
perturbation are the same with the case where shock wave travels from light
to heavy media. The initial physical field is as below:
\begin{equation*}
\left\{
\begin{array}{cc}
(\rho ,u_{x},u_{y},p)_{l}=(3.33333,2.07063,0,7.125)\text{,} &  \\
(\rho ,u_{x},u_{y},p)_{m}=(1,0,0,1)\text{,} &  \\
(\rho ,u_{x},u_{y},p)_{r}=(0.138,0,0,1)\text{,} &
\end{array}%
\right.
\end{equation*}%
The boundary conditions in the $y$-direction and at the left side are the
same as the case where shock wave travels from light to heavy media. Two
different boundaries are applied at the right side: outflow condition (case
I) and reflecting boundary (case II). The computational domain is a
rectangle $0.6\times 0.1$ for case I and $0.3\times 0.1$ for case II, respectively.

Figure 9 shows the simulation results for density field. Figure (a) corresponds
to the outflow boundary and figure (b) corresponds to the reflecting boundary.
Here $\gamma =1.4$. The collision parameters in case I are $s_{5}=10^{4}$, $%
10^{5}$ for the others, and in case II are $s_{5}=10^{3}$, $10^{5}$ for the
others. Simulation results show the following physical procedure: When the
shock wave passes the interface, a reflected rarefaction wave to the left
and a transmission wave to the right are generated. The pressure of heavy
fluid near the crest is greater than the light fluid pressure. Driven by the
pressure gradient, the perturbation amplitude decreases with the interface
motion to the right. Then, the peak and valley of initial interface invert,
the heavy and light fluids gradually penetrate into each other as time goes
on, the light fluid \textquotedblleft rises" to form a bubble and the heavy
fluid \textquotedblleft falls" to generate a spike. In case I, the
transmission wave continues to move to the right, and no longer interacts
with the interface. The disturbance of the interface continues to grow,
eventually forming a mushroom shape. In case II, the transmission wave
reaches the solid wall on the right and reflects to the left, encounters the
interface again. This is known as the \textquotedblleft reshocking" process.
Following reshocking, the interface is compressed, as seen from the kink in
the bubble. Furthermore, the amplitude grows more rapidly than prior to
reshocking, the increased growth is due to the additional vorticity
deposited on the evolving interface during reshocking. The pressure contours
and velocity vectors near the interface at time $t=0.08$ are shown in
Fig.10. Figure 11 shows the change of disturbance amplitudes with time,
corresponding to case I and case II, respectively. Because of the reshocking
effect, a significant difference between Fig.11(a) and Fig.11(b) can be
observed.
\begin{figure}[tbp]
\center\includegraphics*[ width=0.88\textwidth]{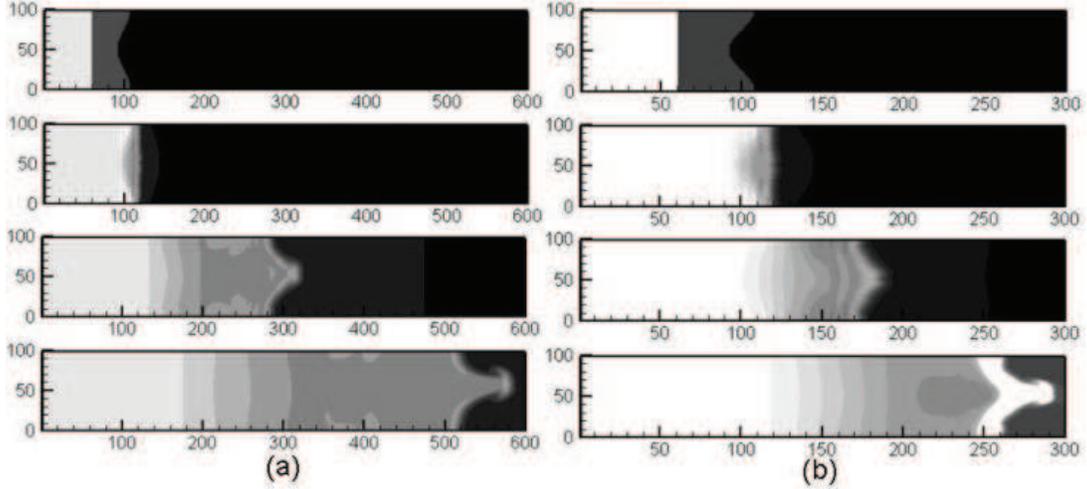}
\caption{Snapshots of RM instability (from heavy to light medium). (a)
Outflow boundary. From top to bottom, $t=0$, $0.02$, $0.08$, $0.16$,
respectively. (b) Reflecting boundary. From top to bottom, $t=0$, $0.02$, $%
0.04$, $0.08$, respectively. From deep to light color, the level corresponds
to the increase of density.}
\end{figure}

\begin{figure}[tbp]
\center\includegraphics*[ width=0.68\textwidth]{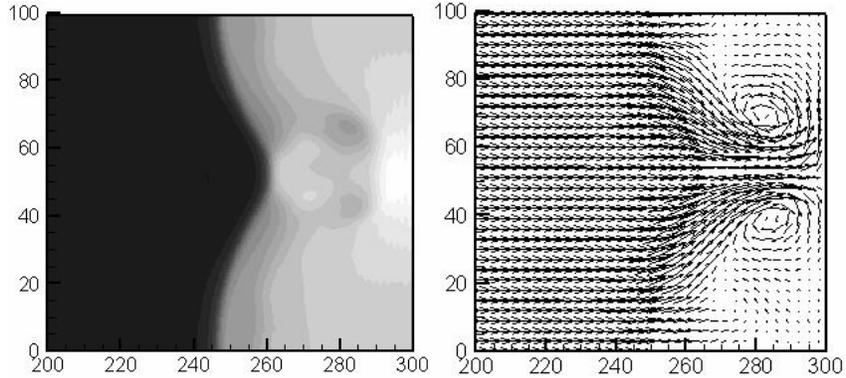}
\caption{Pressure contours and velocity vectors at time $t=0.08$ (from heavy
to light medium, reflecting boundary). From deep to light color, the level
corresponds to the increase of pressure.}
\end{figure}
\begin{figure}[tbp]
\center\includegraphics*
[bbllx=22pt,bblly=79pt,bburx=563pt,bbury=422pt,angle=0,width=0.78\textwidth]
{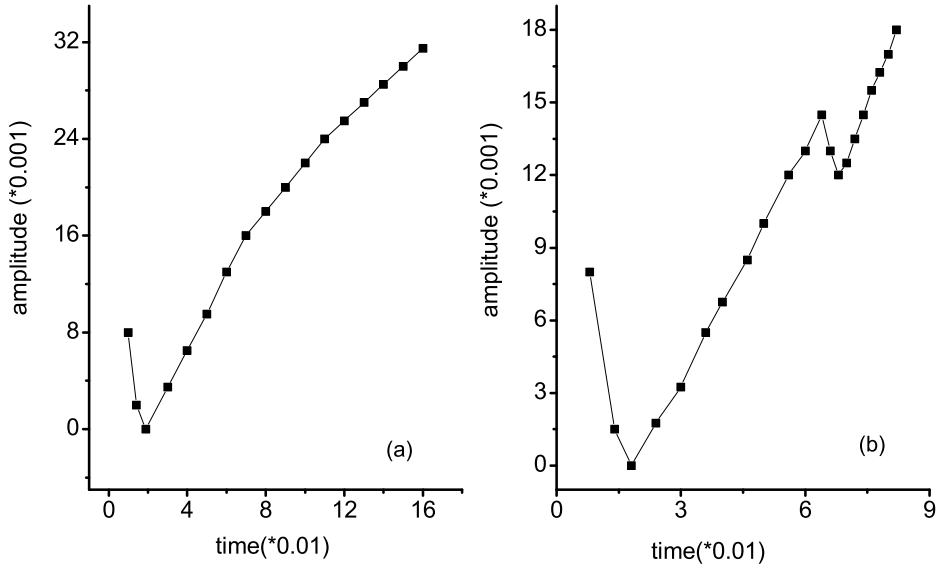}
\caption{Amplitude change with time (from heavy to light medium). (a)
Outflow boundary, (b) Reflecting boundary.}
\end{figure}


The interface reversal phenomenon is observed in the second case. With the
interaction between shock wave and interface, disturbance will grow
continuously. In the early stage, logarithm of growth rate is nearly linear
with time, while changes into the non-linear in the late stage, spikes and
bubbles occur.

\subsection{Kelvin-Helmholtz instability}

During the later stage KH instability strengthens the nonlinear developing
of RT and RM instabilities, enhances the small scale mixing. In some cases,
it may break the spkies. But in some cases, we failed to observe the full
effects of KH instability. For example, in the Eagle Nebula, why has the
famous ``Pillars of Creation" so large scale structures, instead of being
broken by many small scale vortices? There must be some mechanisms to
restrain the KH instability. Therefore, people study the KH instability from
two sides. How does the KH instability evolve? How to enhance or restrain
the KH instability? The strong nonlinearity and multiscale interactions make
difficult theoretical study. The very complex 3D behavior challenge
experimental diagnosis. Our LB modeling and simulation aim to help
understand better the KH instability from both the two sides.

For investigating the Kelvin-Helmholtz instability, we set the following
initial physical field,
\begin{equation}
\rho (x)=\frac{{\rho _{L}+\rho _{R}}}{2}-\frac{{\rho _{L}-\rho _{R}}}{2}%
\tanh (\frac{x}{{D_{\rho }}})\text{,}
\end{equation}%
\begin{equation}
v(x)=\frac{{v_{L}+v_{R}}}{2}-\frac{{v_{L}-v_{R}}}{2}\tanh (\frac{x}{{D_{v}}})%
\text{,}
\end{equation}%
\begin{equation}
P_{L}=P_{R}=P\text{,}
\end{equation}%
where we have two characteristic length scales, ${D_{\rho }}$ and ${D_{v}}$,
which are the widths of density and velocity transition layers,
respectively. ${\rho _{L}=5.0}$ (${\rho _{R}=2.0}$) is the density away from
the interface of the left (right) fluid. ${v_{L}=0.5}$ (${v_{R}=-0.5}$) is
the velocity away from the interface in $y$-direction of the left (right)
fluid, and $P_{L}$ ($P_{R}$)$=2.5$ is the pressure in the left (right) side.
The system can be approximately thought of as ``incompressible". The whole
calculation domain is a rectangle with length $0.6$ and height $0.2$, which
is divided into $600\times 200$ uniform meshes. A simple velocity
perturbation in the $x$-direction is introduced to trigger the KH rollup and
it is in the following form
\begin{equation}
u=u_{0}\sin (ky)\exp (-kx)\text{,}  \label{uu}
\end{equation}%
where $u_{0}=0.02$ is the amplitude of the perturbation. Here, $k$ is the
wave number of the initial perturbation, and is set to be $10\pi $. The time
step is $\Delta t=10^{-5}$.

At the initial linear increasing stage of KH INSTABILITY, the amplitude $\eta$ of
perturbation evolves according to the following relation, $\eta =\eta
_{0}e^{\gamma t}$, where $\gamma$ is the growth coefficient and is dependent
on the gradient of tangential velocity and gradient of density around the
interface. In other words, $\gamma$ is dependent on the width of velocity
transition layer $D_v $ and width of density transition layer $D_{\rho}$. We
discuss separately the KHI in three cases, (i) $D_v$ is variable and $%
D_{\rho}$ is fixed, (ii) $D_{\rho}$ is variable and $D_v$ is fixed, (iii)
both $D_{\rho}$ and $D_{v}$ are variable. The increasing rate $\gamma$ for
cases (i), (ii) and (iii) are referred to as $\gamma_v$, $\gamma_{\rho}$ and
$\gamma_R$, respectively. We numerically obtain $\gamma_v$, $\gamma_{\rho}$
and $\gamma_R$ via fitting the curves of $\ln E_x |_{\max} (t)$ versus the
time $t$, where $E_x |_{\max} (t)$ is the maximum of $E_x (x, y, t)$ in the
whole computational domain, $E_{x} (x,y,t) = \rho (x,y,t) u^2(x,y,t)/2$ is
the perturbed kinetic energy at the position ($x$, $y$) at each time step $t$%
.

Although viscosity damps the evolution of the KH INSTABILITY, here we focus on cases
such as in inertial confined fusion where effects of the viscosity are
generally negligible. Therefore, throughout the simulations, $\tau$ is set
to be $10^{-5}$ to reduce the physical viscosity. Boundary conditions are as
below. Periodic in the $y$-direction and outflow (zero gradient) in the $x$%
-direction.

\subsubsection{Velocity gradient effect}

\begin{figure}[tbp]
\includegraphics*[ width=0.68\textwidth]{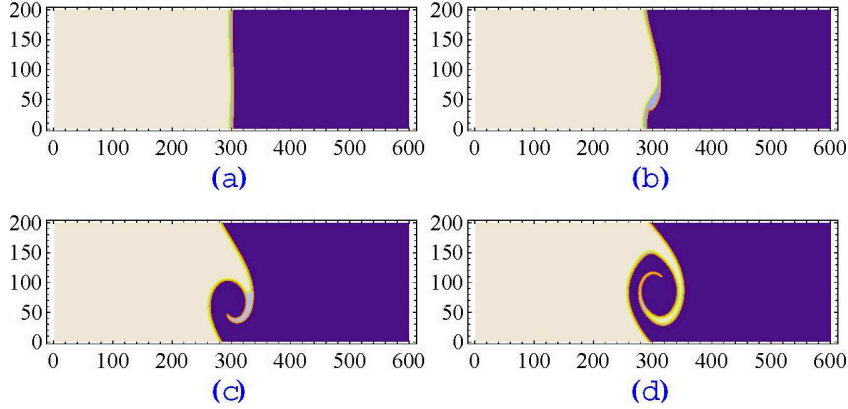}
\caption{(Color online) Density evolutions of KH INSTABILITY simulated using the LB
model, where $D_v = 4$ and ${D_{\protect\rho }=8}$, $t = 0.1$ in (a), $t =
0.3$ in (b), $t = 0.5$ in (c), and $t = 0.7$ in (d). }
\end{figure}

Figure 12 shows the evolution of the density field for the case with ${%
D_{v}=4}$ and ${D_{\rho }=8}$ at four different times. At $t=0.3$ the
interface has been wiggling under the initial perturbation and velocity
shear. A nicely rolled vortex occurs and develops around the initial
interface after the initial linear growth stage. The vortex becomes larger
with time and a mixing layer forms around the initial interface.

\begin{figure}[tbp]
\includegraphics*[bbllx=127pt,bblly=371pt,bburx=501pt,bbury=771pt,angle=0,width=0.68\textwidth]
{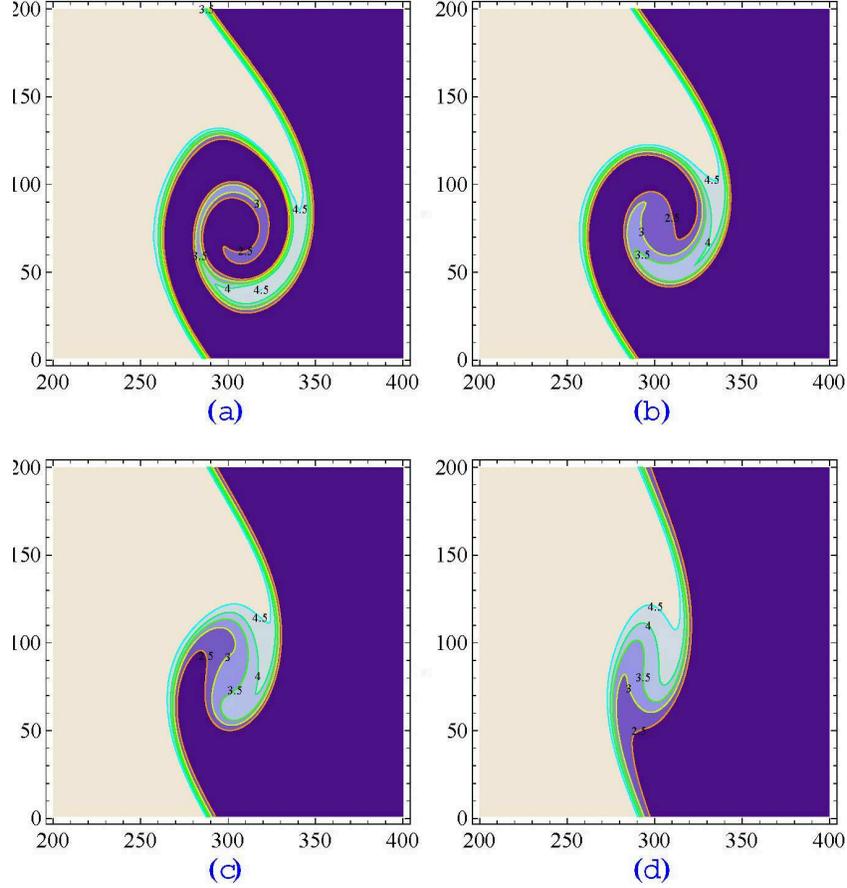}
\caption{(Color online) Vortices in the mixing layer as a function of ${D_{v}%
}$ at $t=0.6$, where ${D_{v}=4}$ in (a), ${D_{v}=8}$ in (b), ${D_{v}=12}$ in
(c), and ${D_{v}=16}$ in (d). The density transition layer ${D_{\protect\rho %
}}$ is fixed to be $8$. }
\end{figure}

To investigate the velocity gradient effect, we fix the width of the density
transition layer. Figure 13 shows the density field for various ${D_{v}}$ at
the same time, where $D_{\rho }=8$, $t=0.6$ and ${D_{v}=4}$, ${8}$, ${12}$, $%
16$ in (a)-(d), respectively. Five contour lines are plotted in each plot.
It is clear that the width of the velocity transition layer significantly
affects the evolution of KH instability. The larger the value of $D_{v}$,
the weaker the KH instability, and the later the vortex appears. In Figs.(a)
and (b), large vortices have been formed demonstrating that the evolution is
embarking on the nonlinear stage. While in Figs. (c) and (d), the evolution
is in the weakly nonlinear stage. Figures (a)-(d) show that a wider velocity
transition zone is helpful for stabilizing the KH instability.

\begin{figure}[tbp]
\includegraphics*[ width=0.68\textwidth]{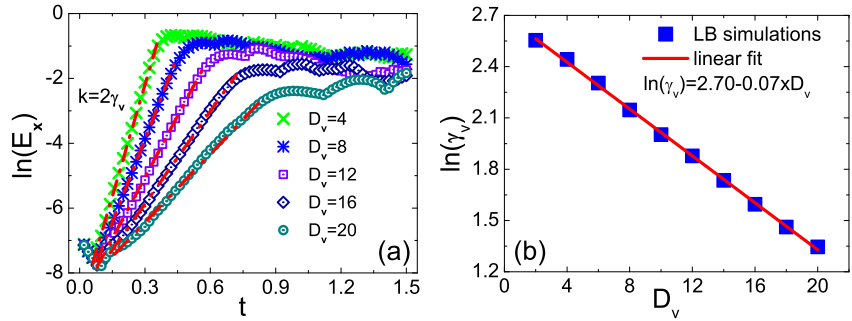}
\caption{(Color online) (a) Time evolution of the perturbed peak kinetic
energy $E_x |_{\max}$ along the $x$-axis in $\ln $-linear scale for various
widths of velocity transition layer. The dash-dotted lines represent the
linear fits to the initial linear growth regimes. (b) Linear growth rate as a
function of the width ${D_{v}}$ of velocity transition layer. }
\end{figure}

The peak kinetic energy $E_x |_{\max}$ partly indicates the interacting
strength of two different fluids. Figure 14(a) shows that logarithm of $%
E_x|_{\max}$ versus time. The initial state shows a linear behavior. The
slope $k$ increases with decreasing the width $D_v$. After the initial
stage, $\ln (E_x|_{max})$ approaches a saturation value via a nonlinear
growth stage. During the initial linear stage, we have $E_{x}\propto
u^{2}\propto \left(e^{\gamma t}\right)^{2}$. So, the slope $k$ here can be
used to calculate the growth coefficient $\gamma$ in the linear growth
stage, $k = 2\gamma$. The logarithm of $\gamma$ decreases linearly with $D_v$
[see Fig.14(b)]. Our LB results confirm the theoretical analysis of Wang, et
al. \cite{wang-pop-2010}. In the classical case, the linear growth rate is $%
\gamma _{c}=k\sqrt{\rho _{1}\rho _{2}}(v_{1}-v_{2})/(\rho _{1}+\rho
_{2})\propto \Delta v$, where $\Delta v$ is the shear velocity difference. A
wider transition layer decreases the local or the effective shear velocity
difference $\Delta {v}$, which results in a smaller linear growth rate and a
longer linear growth time.

\subsubsection{Density gradient effect}

\begin{figure}[tbp]
\includegraphics*[ width=0.68\textwidth]{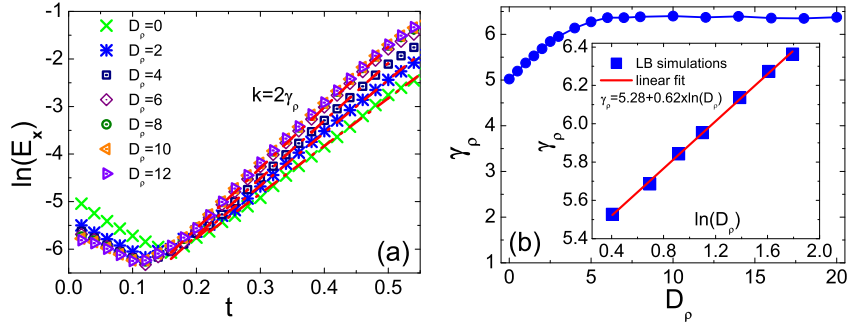}
\caption{(Color online) (a) Time evolution of the logarithm of the peak
kinetic energy $E_x |_{\max}$ along the $x$-axis for various widths of
density transition layers. (b) Linear growth rate as a function of the width
${D}_{\protect\rho }$ of density transition layer. }
\end{figure}

The density gradient effect is investigated in a similar way. Here $D_{v}$
is fixed. The initial conditions are described as, $({\rho _{L}}${, }${v_{L}}
${, }$P_{L})=(5.0$, $0.5$, $1.5)$ and $({\rho _{R}}${, }${v_{R}}${,} $%
P_{R})=(1.25$, $-0.5$, $1.5)$. Figure 15(a) shows evolution of the logarithm
of peak kinetic energy $E_{x}|_{\max }$ along the $x$-axis versus time $t$
for various widths of density transition layers. Here $D_{v}=2$, $\Delta
x=\Delta y=0.002$, $\Delta t=10^{-5}$. Results for $D_{\rho }=0$, $2$, $4$, $%
6$, $8$, $10$, and $12$ are shown. For fixed width of velocity transition
layer and density difference, the linear growth rate first increases with
the width $D_{\rho }$. But when ${D_{\rho }}$ is large than a critical value
which is about $6$, it does not vary significantly any more [see Fig.15(b)
]. During the linear growth stage, $\gamma _{\rho }$ increases linearly with
the logarithm of $D_{\rho }$.Figures 14 and 15 indicate the effective
interaction width of $D_{\rho }$ is less than that of $D_{v}$. The LB
results here confirm also the theoretical analysis of Wang, et al. \cite%
{wang-pop-2010}. In the classical case, the square of the linear growth rate
is $\gamma _{c}^{2}=k^{2}\rho _{1}\rho _{2}(v_{1}-v_{2})^{2}/(\rho _{1}+\rho
_{2})^{2}\propto (1-A^{2})\Delta v^{2}$, where $A=(\rho _{1}-\rho
_{2})/(\rho _{1}+\rho _{2})$ is the Atwood number. A wider density
transition zone reduces the Atwood number around the interface. Then in the
process of exchanging momentum in the direction normal to the interface, the
perturbation can obtain more energy from the shear kinetic energy than in
cases with sharper interfaces. Therefore, a thinner density transition layer
is helpful to restrain the KH instability.

\subsubsection{Hybrid effects of velocity and density gradients}

In practical systems, at the interface of two fluids with a tangential
velocity difference, both the velocity and the density gradients exist.
There is a competition between effects of the two kinds of gradients. We
introduce a coefficient $R=D_{\rho}/D_{v}$ through which we analyze the
combined effects. The linear growth rate versus $D_{\rho}$ under various
values of $R$ is shown in Fig.16. Here $R=0.5$, $1$, $2$, and $5$, as shown
in the legend. On the whole, the hybrid effect of the two kinds of gradients
is to reduce the linear growth rate $\gamma_R$. Only at small $D_{\rho }$
and when $R>1$, the hybrid effect makes larger the linear growth rate. This
indicates again that the effective interaction width of the velocity
transition layer ${D_{v}^{E}}$ is wider than that of density transition
layer ${D_{\rho}^{E}}$.
\begin{figure}[tbp]
\includegraphics*[scale=0.6]{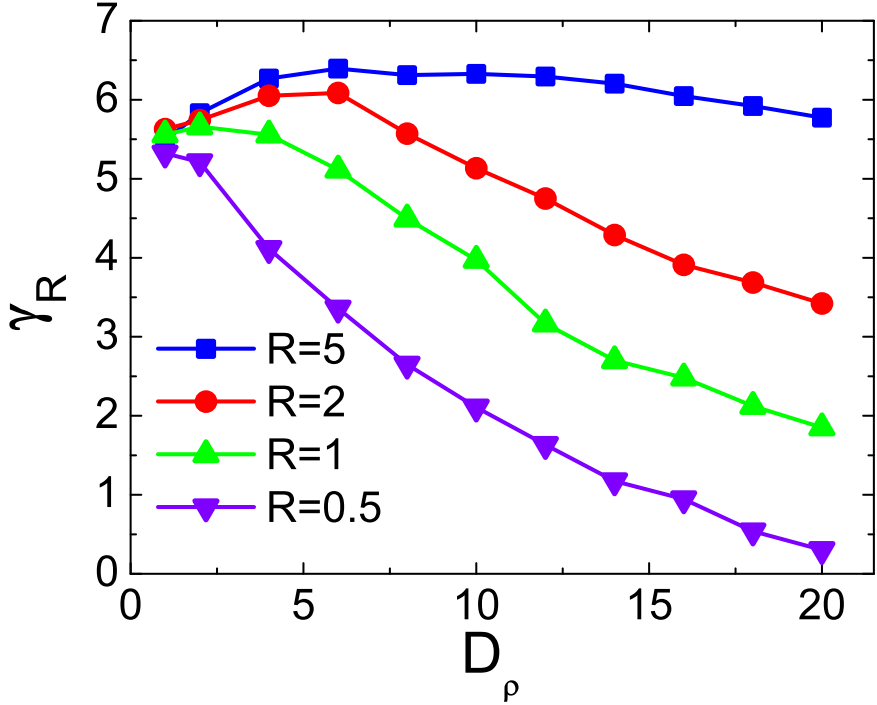}
\caption{(Color online) The linear growth rate versus the width of density
transition layer for $R=0.5$, $1$, $2$, and $5$. The initial density, shear
velocity and pressure of the two fluids are $({\protect\rho _{L}}${, }${v_{L}%
}${, }$P_{L})=(5.0$, $0.5$, $1.5)$ and $({\protect\rho _{R}}${, }${v_{R}}${,
}$P_{R})=(1.25$, $-0.5$, $1.5).$ }
\end{figure}

\section{Conclusions}

Both the LB and the hydrodynamic equations are simplified dynamic models of
practical systems. Compared with the latter, the former puts the physical
modeling on a more fundamental level. When numerically study a physical
procedure, the working dynamic model is not the one evolving continuously in
space and time but the one discretized in the code. Improving the discrete
template and reasonably adding viscosity term are in fact some remedies to the
working dynamic model. Compared with the LB based on BGK approximation, the
MRT-LB introduces a new framework where various physical modes can be
considered separately. The developed SRT-LB and MRT-LB are complementary
from the sides of convenience and applicability. Compared with the
hydrodynamic descriptions, both the SRT-LB and MRT-LB present new
measurements for the deviations of systems from their thermodynamic
equilibria. The LB model is being extended to study the compressibility
effects, effects of shocking and detonation, thermal effects on the hydrodynamic instabilities%
 \cite{LB2KHI2011} and multiphase flows \cite%
{LB2MPhase2011PRE,LB2MPhase2011EPL,LB2MPhase2012FrontPhys}, etc., which are
all-important issues in science and engineering.


\section*{Acknowledgements}

The authors thank Prof. Guoxi Ni for many helpful
discussions. AX and GZ acknowledge support of the Science Foundations of
CAEP [under Grant Nos. 2012B0101014 and 2011A0201002]. AX, GZ, YG and XY acknowledge support of National Natural
Science Foundation of China [under Grant Nos. 11075021, 11171038, 11202003 and 91130020]. YG
acknowledges support of Technology Support Program of LangFang [under Grant
Nos. 2010011030 and 201101118/21/23/24].

\end{document}